\documentclass[10pt]{article}
\usepackage{amsmath}
\usepackage{graphicx}
\usepackage{amsfonts}
\usepackage{amssymb}
\usepackage{epsf}
\usepackage{latexsym}

\def\be{\begin{equation}}
\def\ee{\end{equation}}
\def\bea{\begin{eqnarray}}
\def\eea{\end{eqnarray}}
\def\Rho{\mbox{\Large $\rho$}}

\def\bbeta{\mbox{\boldmath$\beta$}}
\def\bgamma{\mbox{\boldmath$\gamma$}}

\def\ttD{\mbox{\tt D}}

\def\ttP{\mbox{\tt P}}
\def\ttQ{\mbox{\tt Q}}

\def\ttp{\mbox{\tt p}}

\def\ml{\mbox{\scriptsize l}}

\def\mn{\mbox{\scriptsize n}}

\def\NSI{Na\"{\i}ve Schr\"{o}dinger Interpretation }
\def\CPI{Conditional Probabilities Interpretation }

\def\foo{\footnote}
\def\hat{\widehat}

\def\beq{\begin{equation}}
\def\eeq{\end{equation}}
\def\bea{\begin{eqnarray}}
\def\eea{\end{eqnarray}}
\def\pa{\partial}
\def\d{\textrm{d}}

\def\ttD{\mbox{\tt D}}
\def\ttP{\mbox{\tt P}}

\def\ttQ{\mbox{\tt Q}}

\def\tea{\mbox{\tt t}}
\def\stea{\mbox{\scriptsize\tt t}}

\def\5Star{\mbox{\Large$\star$}}

\def\cr{\mbox{\scriptsize{\bf $\mbox{ } \times \mbox{ }$}}}

\def\sumi3{\sum\mbox{}_{\mbox{}_{\mbox{\scriptsize $i$=1}}}^3}

\def\sumj3{\sum\mbox{}_{\mbox{}_{\mbox{\scriptsize $j$=1}}}^3}
\def\sumk3{\sum\mbox{}_{\mbox{}_{\mbox{\scriptsize $k$=1}}}^3}

\def\ma{\mbox{a}}

\def\md{\mbox{d}} 
\def\me{\mbox{e}}

\def\mh{\mbox{h}}
\def\mi{\mbox{i}}

\def\ml{\mbox{l}}

\def\mn{\mbox{n}}

\def\mo{\mbox{o}}
\def\mp{\mbox{p}}

\def\mt{\mbox{t}}
\def\muu{\mbox{u}}

\def\mA{\mbox{A}}

\def\mD{\mbox{D}}

\def\mI{\mbox{I}}

\def\mM{\mbox{M}}
 
\def\mO{\mbox{O}}
\def\mP{\mbox{P}}
\def\mQ{\mbox{Q}}
\def\mR{\mbox{R}}
\def\mS{\mbox{S}}

\def\mX{\mbox{X}}

\def\sa{\mbox{\scriptsize a}}

\def\scc{\mbox{\scriptsize c}}
\def\sd{\mbox{\scriptsize d}}
\def\se{\mbox{\scriptsize e}}
\def\sf{\mbox{\scriptsize f}}
 
\def\sh{\mbox{\scriptsize h}} 
\def\si{\mbox{\scriptsize i}}

\def\sll{\mbox{\scriptsize l}}  
\def\sm{\mbox{\scriptsize m}}
\def\sn{\mbox{\scriptsize n}} 
\def\so{\mbox{\scriptsize o}}

\def\sr{\mbox{\scriptsize r}}

\def\st{\mbox{\scriptsize t}}
\def\su{\mbox{\scriptsize u}}

\def\sw{\mbox{\scriptsize w}}

\def\sB{\mbox{\scriptsize B}}

\def\sG{\mbox{\scriptsize G}}
\def\sH{\mbox{\scriptsize H}}

\def\sJ{\mbox{\scriptsize J}}
\def\sK{\mbox{\scriptsize K}}
 
\def\sM{\mbox{\scriptsize M}} 
\def\sN{\mbox{\scriptsize N}} 

\def\sP{\mbox{\scriptsize P}} 
 
\def\sR{\mbox{\scriptsize R}}
\def\sS{\mbox{\scriptsize S}}

\def\sW{\mbox{\scriptsize W}}

\def\eph{\mbox{\scriptsize eph}}
 
\def\eph(B){\mbox{\scriptsize em(JBB)}}

\def\ttP{\mbox{\tt P}}

\def\eph(B){\mbox{\scriptsize emergent(JBB)}}

\def\ta{\mbox{\tiny a}}

\def\td{\mbox{\tiny d}}
\def\te{\mbox{\tiny e}}

\def\th{\mbox{\tiny h}}
\def\ti{\mbox{\tiny i}}

\def\tn{\mbox{\tiny n}}

\def\ttt{\mbox{\tiny t}}

\def\tR{\mbox{\tiny R}}

\def\bg{\mbox{{\bf g}}}
\def\bM{\mbox{{\bf M}}}
\def\bM{\mbox{{\bf M}}}
\def\bN{\mbox{{\bf N}}}

\def\bh{\mbox{{\bf h}}}
\def\bm{\mbox{{\bf m}}}

\def\bq{\mbox{{\bf q}}}

\def\bx{\mbox{{\bf x}}}

\def\sbm{\mbox{{\bf \scriptsize m}}}

\def\fH{\mbox{\sffamily H}}

\def\fW{\mbox{\sffamily W}}

\def\sfA{\mbox{\sffamily{\scriptsize A}}}
\def\sfB{\mbox{\sffamily{\scriptsize B}}}

\def\sfX{\mbox{\sffamily{\scriptsize X}}}

\def\sfZ{\mbox{\sffamily{\scriptsize Z}}}
\def\K{Kucha\v{r} }

\def\bp{\mbox{\bf p}}

\def\bM{\mbox{\bf M}}
\def\bQ{\mbox{\bf Q}}

\begin{document}
\begin{titlepage}
\vspace{.7in}
\begin{center}
\LARGE{\bf PROBLEM OF TIME IN QUANTUM GRAVITY}\normalsize 

\vspace{.1in}

\normalsize

\vspace{.4in}
 
{\large \bf Edward Anderson$^*$}

\vspace{.2in}

\large {\em APC AstroParticule et Cosmologie, Universit\'{e} Paris Diderot CNRS/IN2P3, CEA/Irfu, 
Observatoire de Paris, Sorbonne Paris Cit\'{e}, 10 rue Alice Domon et L\'{e}onie Duquet, 75205 Paris Cedex 13, France} \normalsize

\vspace{.2in}

\large and {\em DAMTP, Centre for Mathematical Sciences, Wilberforce Road, Cambridge CB3 OWA, U.K.} \normalsize

\end{center}

\begin{abstract}

The Problem of Time occurs because the `time' of GR and of ordinary Quantum Theory are mutually incompatible notions.
This is problematic in trying to replace these two branches of physics with a single framework in situations in which the 
conditions of both apply, e.g. in black holes or in the very early universe.
Emphasis in this Review is on the Problem of Time being multi-faceted and on the nature of each of the eight principal facets.  
Namely, the Frozen Formalism Problem, Configurational Relationalism Problem (formerly Sandwich Problem), Foliation Dependence Problem, 
Constraint Closure Problem (formerly Functional Evolution Problem), Multiple Choice Problem, Global Problem of Time, Problem of Beables 
(alias Problem of Observables) and Spacetime Reconstruction/Replacement Problem. 
Strategizing in this Review is not just centred about the Frozen Formalism Problem facet, but rather about each of the eight facets. 
Particular emphasis is placed upon A) relationalism as an underpinning of the facets and as a selector of particular strategies 
(especially a modification of Barbour relationalism, though also with some consideration of Rovelli relationalism).
B) Classifying approaches by the full ordering in which they embrace constrain, quantize, find time/history and find observables, rather 
than only by partial orderings such as ``Dirac-quantize".  
C) Foliation (in)dependence and Spacetime Reconstruction for a wide range of physical theories, strategizing centred about the 
Problem of Beables, the Patching Approach to the Global Problem of Time, and the role of the question-types considered in physics. 
D) The Halliwell- and Gambini--Porto--Pullin-type combined Strategies in the context of semiclassical quantum cosmology.

\end{abstract}

\vspace{3in}

\mbox{ }

\mbox{ } 

\mbox{ }

\mbox{ }  
 
\noindent $^*$ eanderso@apc.univ-paris7.fr 

\end{titlepage}

\section{Introduction}\label{Intro}

The notorious {\bf Problem of Time (POT)} \cite{DeWitt67, Battelle, Kuchar81, Kuchar91, Kuchar99, Kuchar92, I93, Kuchar93, Rovellibook, Kieferbook, 
Smolin08, APOT, FileR} occurs because the `time' of General Relativity (GR) and the `time' of ordinary Quantum Theory are mutually incompatible notions.
This incompatibility leads to a number of problems with trying to replace these two branches of physics with a single framework in situations in 
which the conditions of both apply, such as in black holes or in the very early universe.  
I begin with an outline of some relevant aspects of what time means in general \cite{Jammerteneity, FileR} (Sec 1.1) and for each of these theories (Sec 1.2-3).

\subsection{Some properties often ascribed to time} 

\noindent i) {\bf Time as an ordering}.  
Simultaneity enters here, i.e. a notion of present that separates future and past notions, each of which is perceived differently 
(e.g. one remembers the past but not the future).  
The structure of the present {\it instant} is the whole-universe configuration that includes one's theory's notion of space 
as well as whatever internal spaces along the lines of gauge theory.

\noindent ii) {\bf Causation}: that one phenomenon brings at an earlier value of `the time' brings about another at a later value.  

\noindent iii) {\bf Temporal logic}: this extends more basic (atemporal) logic with ``and then" and ``at time $t_1$" constructs.

\noindent iv) In dynamics, one encounters the idea of {\bf change in time}. 
Thus here time is taken to be a type of container: a parameter with respect to which change is manifest.   
Newtonian absolute time is an example of this (and external to the system itself and continuous).  
Contrast the Leibniz--Mach view of time in Sec 1.3!
There is a long-standing philosophical fork between considering time to be fundamental and time being something that should be eliminated from one's conceptualization of the world. 
Approaches of the second sort are to reduce questions about `{\bf being at a time}' and `{\bf becoming}' to just questions of `{\bf being}' (see Sec \ref{TNE}). 

\noindent v) Mathematically, time is often taken to be modelled by the real line or an interval of this or a discrete approximation of this. 
Though time can easily be {\sl position-dependent} in field theories: $\mt(\bx)$ in place of $t$.
[I use $\tea$ to denote either, and follow suit with other quantities using these fonts for finite theories, field theories and 
portmanteaux of the two. 
This notation of straight symbol for field-theoretic version and slanty symbol for finite theory version shall pervade this article.]  

\noindent vi) Time is additionally habitually taken to be {\bf monotonic} (rather than direction-reversing).   
This makes sense as regards its possessing ordering and causation properties.    
    
\noindent vii) There is to be freedom in prescribing a timefunction as to the {\bf choice of calendar year zero} and {\bf of tick-duration}, 
i.e. if $\tea$ is a timefunction, so is $A + B\,\tea$ for $A$, $B$ constants.

\noindent viii) A good timefunction is {\bf globally valid} \cite{Kuchar92, Hartle96} both over time (antagonist to the half-finite and finite 
interval times unless there is a good physical reason for this) and over space (made necessary by field theories and, to a greater extent generic curved space/GR).   

\noindent ix) It also makes sense to require one's time function to be {\bf operationally meaningful} (computable from observable quantities -- 
tangible and accessible in practise).  

\mbox{ } 

\noindent {\bf Candidate times}.   Simply calling an entity a time or a clock does not make it one.
One needs to check a fairly extensive property list before one can be satisfied that it is a) indeed a time or a clock and b) 
that it is useful for accurate work in comparing theory and experiment.  
However, what is a sufficient set of properties for a candidate time to be a time is disputed because GR and ordinary QM, so this 
a choice of a sufficiency of properties rather than just a check-list.

\subsection{Time in QM and in GR }

\noindent {\bf Newtonian Absolute Time} is what ordinary QM is based upon.
This is a {\sl fixed} background parameter.
Thus time comes to enter ordinary QM as a background parameter that is 

\noindent 1) used to mark the evolution of the system. 

\noindent 2) It is represented  by an {\sl anti}-hermitian operator (unlike the representation of other quantities). 

\noindent 3) The time--energy uncertainty relation $\Delta t^{\sN\se\sw\st\so\sn}\Delta E \geq \hbar/2$ is also given an entirely different meaning 
to that of the other uncertainty relations.  
[I subsequently use $\hbar = c = G = 1$ units.]  

\noindent 4) There is unitary evolution in time, i.e. that probabilities always sum to one.
This evolution is in accord with the theory's time-dependent wave equation (for now a time-dependent Schr\"{o}dinger equation).  
The {\it scalar product} on the Hilbert space of states leads to conserved probability currents.

\noindent 5) Moreover, there is a second dynamical process: collapse of the wavefunction that is held to occur in ordinary QM despite its not being 
described by the evolution equation of the theory.  

\noindent {\bf Special Relativity (SR)} brings in the further notions of 

\noindent 1) a proper time corresponding to each observer, and of 

\noindent 2) time as another coordinate on (for now, flat) spacetime, as opposed to the external absolute time of Newtonian physics.  

\noindent 3) However, time in SR is also external and absolute in the sense of there being (via the existence of Killing vectors) a presupposed set of privileged inertial frames. 
[The quantum theory can be made independent of a choice of frame if it carries a unitary representation of the Poincar\'e group.]    
In that sense in SR all one has done is trade one kind of absolute time for another \cite{Kiefer09bis}. 
(The changes are in the notion 
of simultaneity \cite{Jammerteneity} rather than in removal of the absolute; also the type of wave equation is now e.g. Klein--Gordon or Dirac.)
Thus the passage from Newtonian Mechanics to SR, and thus from 

\noindent ordinary QM to relativistic QFT does not greatly affect the role of time.  

\noindent It is only GR that frees one from absolute time, and so that the SR to GR conceptual leap is in many ways bigger than the 
one from Newtonian Mechanics to SR.  
Here, 

\noindent 1) the conventional {\bf GR spacetime formulation} is in terms of $({\mM}, \bg)$ for ${\mM}$ the spacetime topological manifold and $\bg$ 
\noindent the indefinite spacetime 4-metric that obeys the Einstein field equations. 
[I usually restrict to the vacuum case for
simplicity, but extending this article's study to include the ordinary matter fields is 
unproblematic.]  

\noindent 2) Time is a {\sl general} spacetime coordinate in GR, clashing with ordinary QM's having held time to be a unique and sui generis extraneous quantity.   

\noindent 3) {\bf The generic spacetime has no timelike Killing vector}, so much of the preceding QM structure ceases to have an analogue.  
This puts an end to being able to pass between absolute structures in going from Newtonian Mechanics to SR, since the latter's notion of absolute time is 
based on Minkowski spacetime's timelike Killing vector.
Now instead one has to pass from having privileged frame classes to dealing with the spacetime diffeomorphisms (which are a much larger and less 
trivial group of transformations).

\noindent 4) GR has the time-ordering property, and the notion of causality as an extension of SR but now with matter and gravitation 
influencing the larger-scale causal properties.  

\noindent 5) GR additionally has a time non-orientability notion \cite{Wald}, and a closed timelike curve notion; both of these are usually 
held to be undesirable features for a physical solution to possess.

\noindent 6) GR can furthermore be formulated as an evolution of spatial geometry -- the spatial geometries themselves being 
the entities (configurations) that undergo the GR theory's dynamics: GR is a {\sl Geometrodynamics} (see Sec 2 for more).  

\noindent 7) Moreover, time as a general choice of coordinate is embodied geometrically in how GR spacetime is sliced into a 
{\bf sequence of} (or {\bf foliation by}) {\bf spacelike hypersurfaces} corresponding to constant values of that time.  
There is one timefunction per choice of foliation; thus time in GR is said to be `{\bf many-fingered}' (see Secs 2 and \ref{Facets} 
for further detail).  
\noindent That space and time can be thought of as different and yet spacetime also carries insights probably accounts for why 
a number of POT facets are already present in classical GR (see Secs 3 and \ref{Facets}).

\mbox{ } 

\noindent External time is furthermore incompatible with describing truly {\sl closed} systems such as closed-universe Quantum Cosmology itself. 
In this setting, Page and Wootters \cite{PW83} have argued convincingly that such a system's only physical states are {\sl eigenstates} 
of the Hamiltonian operator, which have an essentially trivial time evolution. 

\mbox{ }

\noindent Furthermore, the idea of events happening at a single time plays a crucial role in the technical and conceptual foundations 
of QM, as follows \cite{I93}.  

\noindent 1) {\sl Measurements} made at a particular time are a fundamental ingredient of the conventional Copenhagen 
interpretation (which is anchored on the existence of a privileged time).  
Then in particular \cite{I93}, an {\bf observable} is a quantity whose value can be measured at a `given time'.  
On the other hand, a `history' has no direct physical meaning except in so far as it refers to the outcome of a sequence of time-ordered measurements.
That histories are part QM-like in path integral form, but also capable of harbouring GR-like features, may bode well for (part of) a reconciliation of GR and QM.  
\noindent How QM should be {\sl interpreted} for the universe as a whole is then a recurring theme (see e.g. Secs 6.3--4).  

\noindent 2) In constructing a Hilbert space for a quantum theory, one is to select a complete set of observables that are required to 
{\sl commute} at a fixed value of time -- i.e. obey {\bf equal-time commutation relations}.
\noindent This again extends to SR by considering not one Newtonian time but the set of relativistic inertial frames; 
then one's QM can be made independent of frame choice via it carrying a unitary representation of the Poincar\'e group. 
For a relativistic QFT, the above equal-time commutation relations issue is closely related to requiring {\it microcausality} \cite{I93}: 
\beq
[\widehat{\phi}(X^{\mu}),\widehat{\phi}(Y^{\mu})]    = 0     
\label{microcausal}
\eeq
for relativistic quantum field operators $\hat{\psi}$ and all spacelike-separated spacetime points $X^{\mu}$ and $Y^{\mu}$.  
However, observables and some notion of equal-time commutation relations pose significant difficulties in the context of GR.

\subsection{Background Independence, relational criteria and the POT}

Our attitude is that Background Independence is philosophically-desirable and a lesson to be taken on board from GR, quite separately from any detail 
of the form of its field equations that relativize gravitation.
`Background Independence' is to be taken here in the sense of freedom from absolute structures, according to the following.

\mbox{ } 

\noindent {\bf Temporal relationalism} \cite{BB82, B94I, GrybJac, FileR} concerns time being not primary for the universe as a whole (Leibniz's view).
This is to be implemented by reparametrization-invariance in the absense of extraneous time-like variables 
The conundrum of our apparent local experience of time is then to be resolved along the lines of Mach's time is to be abstracted from change' 
(see Secs 2 and 3 for more detail of these last two sentences).  

\noindent {\bf Configurational relationalism} consists in regarding the configuration space $Q$ of dynamical objects to possess a group $G$ of 
transformations that are physically irrelevant \cite{BB82, RWR, Kieferbook, GrybThesis, ARel}.  
A classic example are the translations and rotations with respect to absolute space, though the internal-space transformations of Gauge Theory are 
also of this nature \cite{FileR}.

\mbox{ }  

\noindent Moreover, Background Independence already implies a number of POT facets at the classical level, and the rest at the
quantum level.  
So whenever one adheres to Background Independence in the above sense, some particular form of the POT 
arises as a direct consequence, and then one must face it.  

\mbox{ } 

\noindent Note 1) Whilst Einstein was inspired by Mach, he did not implement Mach directly in setting up GR.  
However, direct implementation of Mach via temporal and Configurational Relationalism does give back (\cite{RWR}, Secs 3 and 7) the dynamical formulation of GR.

\noindent Note 2) Background Independence criteria are held to be an important feature in e.g. Geometrodynamics and Loop Quantum Gravity (LQG).

\noindent Note 3) Perturbative covariant quantization involves treating the metric as a small perturbation about a fixed background metric.  
This is neither background-independent nor successful on its own terms, at least in the many steps of this program that have been completed, 
due to nonrenormalizability, and non-unitarity in higher-curvature cases.
Perturbative String Theory is another background-dependent approach.  
This amounts to taking time to be a fixed-background SR-like notion rather than a GR-like one at the deepest level of this formulation.  
Here, GR's field equations are emergent, so that problems associated with them are not held to be fundamental but should be referred further 
along to background spacetime metric structure that the strings move in.  
Moreover, technical issues then drive one to seek nonperturbative background-independent strategies and then POT issues resurface among the 
various possibilities for the background-independent nature of M-theory. 

\noindent Note 4) LQG has a greater degree of Background Independence than perturbative String Theory: it is independent of background {\sl metric} structure. 
[However the argument for this can be repeated at the level of background {\sl topological structure}, for which one is presently very largely  unprepared.]

\subsection{Outline of the rest of this Review}

\noindent Whilst I consider the Frozen Formalism Problem (FFP) in Sec 3 and a number of conceptualizations that have been suggested to solve it in Sec 6, 
unlike in previous POT reviews \cite{Kuchar81, Kuchar91, Kuchar93, Kuchar99, Kieferbook, APOT}, the present review concentrates on how there are other 
facets (Secs 4 and 5), and on freeing conceptual strategizing from dealing primarily with the FFP (Secs 7 to 13).  
The seven other facets are \cite{Kuchar92, I93} Configurational Relationalism, that generalizes the Best Matching Problem that itself generalizes 
the Thin Sandwich Problem \cite{W63}, The Problem of Beables (alias Problem of Observables), the Foliation Dependence Problem, 
the Constraints Closure Problem, the Global POT, the Multiple Choice Problem and the Spacetime Reconstruction/Replacement Problem.
Secs 6 to 13 additionally update the classic reviews \cite{Kuchar92, I93} due to the almost 20 years of work since.  
\noindent I use relational/Machian/background-independent criteria as a characterization/possible deeper explanation of POT facets, and then as a 
selection criterion on theories and on POT strategies.
These select for semiclassical, many timeless approaches and histories approaches.
I end by considering composite of these strategies in Sec 14: Halliwell-type \cite{HT02, H03, HW06, H09, H11, AHall} and Gambini--Porto--Pullin-type \cite{PGP1, GPPT} approaches.

\section{Dynamical formulation of GR}\label{DynGR}

In dynamical/canonical approaches, the spacetime manifold $\mM$ is `3 + 1' split (into 3-$d$ space and 1 time dimension) into $\Sigma \times \mI$.  
This assumes that the spatial topology $\Sigma$ does not change with coordinate time $t \in \mI  \,\,\underline{\subset} \,\, \mathbb{R}$. 
This furthermore assumes that GR is globally hyperbolic \cite{Wald}, which amounts to determinability of GR evolution from GR initial data; this excludes 
e.g. time non-orientability and closed timelike curves.  
Some considerations involve a single $\Sigma$ embedded into GR spacetime (or without that assumption, usually leading to its deduction, c.f. Sec \ref{Facets}).

Some further considerations require a {\it foliation} ${\cal F} = \{{\cal L}_{\sfA}\}_{\sfA \in A}$, which is a decomposition of an $m$-dimensional manifold 
${\cal M}$ into a disjoint union of connected $p$-dimensional subsets (the {\it leaves} ${\cal L}_{\sfA}$ of the foliation) such that the following holds. 
$m \in {\cal M}$ has a neighbourhood $U$ in which coordinates ($x^1, ..., x^m$): $U \rightarrow \mathbb{R}^m$ such that for each leaf ${\cal L}_{\sfA}$ 
the components of $U \bigcap {\cal L}_{\sfA}$ are described by $x^{p + 1}$ to $x^m$ constant [see Fig 1a)].
The codimension of the foliation is then $c = m - p$. 
For 3 + 1 GR, ${\cal M} = \mM$, so $m = 4$, the leaves ${\cal L}_{\sfA}$ are 3-space hypersurfaces so $p = 3$ and thus $c = 1$: the time dimension.  

{            \begin{figure}[ht]
\centering
\includegraphics[width=0.6\textwidth]{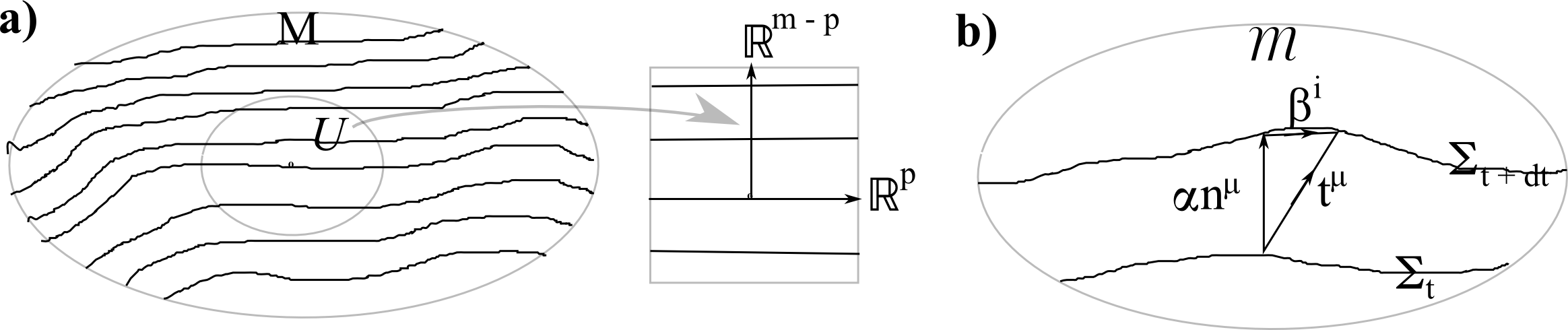}
\caption[Text der im Bilderverzeichnis auftaucht]{        \footnotesize{a) Arnowitt--Deser--Misner 3 + 1 split of $({\mM}, \bg)$. 
b) Picture supporting the text's definition of foliation.}  }
\label{Fig-1}\end{figure}          }

\noindent Given 2 neighbouring hypersurfaces, one can \{3 + 1\}-decompose the spacetime metric $\bg$ between these 
following Arnowitt--Deser--Misner \cite{ADM}  [see Fig \ref{Fig-1}b] into
\beq
g_{\mu\nu} =
\left(
\stackrel{    \mbox{$ \beta_{k}\beta^{k} - \alpha^2$}    }{ \mbox{ }  \mbox{ }  \beta_{i}    }
\stackrel{    \mbox{$\beta_{j}$}    }{  \mbox{ } \mbox{ }  h_{ij}    }
\right)
\mbox{ }  . 
\eeq
Here, $\alpha$ is the lapse (time elapsed), $\beta^{i}$ is the shift (in spatial coordinates between the two slices) and $\bh$ is the intrinsic 
metric on the original hypersurface $\Sigma_t$ itself.
The relation between the foliation vector $n^{\mu}$ and the above \{3 + 1\}-split quantities 
is $t^{\mu} = \alpha n^{\mu} + \beta^{\mu}$, and so $n^{\mu} = \alpha^{-1}[1, - \bbeta]$.   

\noindent Next,\foo{The spatial 
topology $\Sigma$ is taken to be compact without boundary. 
$h_{ij}$ is a spatial 3-metric thereupon, with determinant $h$, covariant derivative $D_{\mu}$, Ricci scalar Ric($\bx;{\bh}$] and conjugate momentum 
$\pi^{ij}$.  
$\Lambda$ is the cosmological constant.
Here, the GR configuration space metric is 

\noindent ${\cal G}^{ijkl} = h^{ik}h^{jl} - h^{ij}h^{kl}$ is the undensitized inverse DeWitt supermetric with determinant ${\cal G}$ and 
inverse ${\cal F}_{ijkl}$ itself the undensitized DeWitt supermetric $h_{ik}h_{jl}  -  h_{ij}h_{kl}/2$. 
The Ricci scalar built out of this metric is $\mbox{Ric}_{\mbox{\scriptsize\boldmath${\cal G}$}}(\bx; \bh]$.
In this article, $\lfloor \mbox{ } \mbox{ } \rfloor$ is a portmanteau of 1) function dependence ( \mbox{ } ) in finite theories and 2) 
functional dependence [ \mbox{ } ] or mixed function-functional dependence ( \mbox{ } ; \mbox{ } ] in field theories.
$\pounds_V$ is the Lie derivative with respect to a vector field $V^i$. 
(4) superscripts denote spacetime tensors; $G^{(4)}_{\mu\nu}$ is the spacetime Einstein tensor.}
the {\it extrinsic curvature} of a hypersurface is the rate of change of the normal along that hypersurface and thus of the bending of that hypersurface relative to its ambient space: 
\beq
K_{\mu\nu} := h_{\mu}\mbox{}^{\rho}\nabla_{\rho}n_{\nu} = \pounds_{n}h_{\mu\nu}/2 \mbox{ } ,
\eeq
or (since it is a hypersurface object so it can be thought of as a spatial tensor as well as a spacetime one),  
\beq
K_{ij} =  \{\dot{h}_{ij} - \pounds_{\beta}h_{ij}\}/{2\alpha} = \{\dot{h}_{ij} - 2D_{(i}\beta_{j)}\}/{2\alpha}  \mbox{ } .
\eeq
As well as this being an important characterizer of hypersurfaces, it is relevant due to its close connection to the GR momenta, 
\beq
\pi^{ij} = - \sqrt{h}\{ K^{ij} - K h^{ij} \} \mbox{ } , 
\eeq
for $K := K_{ij}h^{ij}$.
Next, the GR constraints arise from the 3 + 1 split of the (here vacuum) Einstein field equations, which have the given forms in terms of each of 
$K_{ij}$ and of $\pi^{ij}$, 
\beq
\mbox{Hamiltonian constraint, } {\cal H} := 2{G}^{(4)}_{\perp\perp} = K^2 - K_{ij} K^{ij}  + \mbox{Ric}(\bx; \bh]  = 
{\cal F}_{ijkl}\pi^{ij}\pi^{kl}/\sqrt{h} - \sqrt{h}\mbox{Ric}(\bx; \bh]  = 0  \mbox{ } ,   
\eeq
\be
\mbox{momentum constraint, } {\cal H}_{i} :=  2{G}^{(4)}_{i\perp} = 2\{D_{j}{K^{j}}_{i} - D_{i }K\}  = -2D_{j}\pi^{j}\mbox{}_{i} = 0 \mbox{ } .  
\label{Mom}
\eeq
Note 1) The $K_{ij}$-forms of these serve to identify these as contractions of the Gauss--Codazzi equations for the embedding of spatial 
3-slice into spacetime (a higher-$d$ indefinite-signature generalization of Gauss' Theorema Egregium).

\noindent Note 2) The GR momentum constraint is straightforwardly interpretable in terms of the physicality residing solely in terms of 
the 3-geometry of space rather than in its coordinatization/point-identification.  
However, interpreting the GR Hamiltonian constraint is tougher; this leads to the FFP. 
 
\noindent Note 3) On the other hand, the GR momentum constraint often becomes entwined in technical problems that afflict POT strategies.  
\noindent Some of these issues further involve the specific 3-diffeomorphism Diff($\Sigma$) nature of the associated group of 
physically-irrelevant transformations [and by extension Superspace($\Sigma$) = Riem($\Sigma$)/Diff($\Sigma$), for Riem($\Sigma$) the configuration 
space of spatial 3-metrics on a fixed $\Sigma$].
There are various further ways in which the spacetime diffeomorphisms Diff(${\mM}$) and their hypersurface-split version enter various 
POT Facets and Strategies; these are the subject of Secs \ref{CR} and 9.  

\mbox{ } 

Whilst the first two of these are just infinite-dimensional Lie groups, the third corresponds to the `Dirac Algebra'\footnote{Computationally, this 
follows most succinctly from the Bianchi identity \cite{Wald} using the first forms of eqs (6,7).}  
of the GR constraints,   
\be 
\{    \mbox{\bf{${\cal H}$}}(\bbeta^{\prime}), \mbox{\bf{${\cal H}$}}(\bbeta)    \} = \pounds_{\beta}\mbox{\bf{${\cal H}$}}(\bbeta^{\prime}) 
\mbox{ } , 
\label{mommom} 
\ee 
\be 
\{    {\cal H}(\alpha), \mbox{\bf{${\cal H}$}}(\bbeta)\} = 
\pounds_{\beta}{\cal H}(\alpha) 
\mbox{ } , 
\label{hammom} 
\ee 
\be 
\{{\cal H}(\alpha), {\cal H}(\alpha^{\prime})\} = \mbox{\bf{${\cal H}$}}(\bgamma) 
\mbox{ }\mbox{ for  } \mbox{ } 
\gamma^{i} := hh^{ij}\{\alpha\pa_{j} \alpha^{\prime} - \alpha^{\prime}\pa_{j}\alpha\} 
\mbox{ } .  
\label{hamham} 
\ee 
Since the third bracket gives {\sl structure functions}, this is far more complicated than a Lie algebra (it is a {\it Lie algebroid} \cite{BojoBook}).  
(\ref{mommom}) is the closure of the Lie algebra of 3-diffeomorphisms on the 3-surface, whilst (\ref{hammom}) just means that ${\cal H}$ 
is a scalar density; neither of these have any dynamical content.
It is the third bracket then that has dynamical content.
The Dirac algebroid has remarkable properties at the classical level, as exposited in Sec \ref{Facets}.  
However, that the Dirac algebroid of the classical GR constraints is not a Lie algebra does limit many a quantization approach \cite{I84}.  

The Diff(${\cal M}$) that are not among the Diff($\Sigma$) do cause a number of serious difficulties with quantization schemes \cite{I84}.  

\noindent The quantum GR Hamiltonian constraint is a Wheeler--DeWitt equation (WDE),\footnote{The inverted commas indicate that the WDE  
has, in addition to the POT, various technical problems, including regularization problems (not at all straightforward in 
background-independent theories), what meaning to ascribe for functional differential equations, and operator-ordering issues (some of which are parametrized by the number $\xi$).}
\beq
\hat{\cal H}\Psi := -\hbar^2 \mbox{`}\left\{\triangle_{\mbox{\scriptsize\boldmath${\cal G}$}} 
- \xi \,\mbox{Ric}_{\mbox{\scriptsize\boldmath${\cal G}$}}(\bx; \bh]\right\}\mbox{'}\,\Psi -  \sqrt{h}\mbox{Ric}(\bx; \bh]\Psi   = 0  \label{WDE3} \mbox{ }  ,      \mbox{ } \mbox{ for } 
\triangle_{\mbox{\scriptsize\boldmath${\cal G}$}} := \frac{1}{\sqrt{{\cal G}}}\frac{\delta}{\delta h^{{ij}}}
\left\{
\sqrt{{\cal G}}{\cal F}^{ijkl}\frac{\delta\Psi}{\delta h^{{kl}}}
\right\}
\eeq
\beq
\widehat{\cal H}_{i}\Psi := -2D_{j}h_{ik}\delta\Psi/\delta h_{jk} = 0 \mbox{ } .
\eeq
\noindent In the {\bf Ashtekar variables} formulation of GR \cite{Ash91}, qualitative details of the POT facets remain largely unchanged.
\noindent However, using these is relevant to some of the strategies subsequently discussed, so I provide an outline. 
\noindent Pass from $h_{ij}$, $\pi^{ij}$ to a $SU(2)$ connection ${\mathbb{A}_{i}}^{\mbox{\scriptsize\tt AB\normalfont\normalsize}}$ 
and its conjugate momentum ${\mathbb{E}^{i}}_{\mbox{\scriptsize\tt AB\normalfont\normalsize}}$ 
[which is related to the 3-metric by $h_{ij} = - \mbox{tr}(\mathbb{E}_{i}\mathbb{E}_{j})$].\foo{The 
capital typewriter indices denote the Ashtekar variable use of spinorial $SU(2)$ indices.  
tr denotes the trace over these. $\mathbb{D}_i$ is the $SU(2)$ covariant derivative as defined in the first equality of (\ref{ashgauss}). 
$|[\mbox{ }, \mbox{ } ]|$ denotes the classical Yang--Mills-type commutator.}  
%
This formulation's constraints are

\noindent 
\be
\mathbb{D}_{i}  
{\mathbb{E}^{i}}_{\mbox{\scriptsize\tt AB\normalfont\normalsize}} 
\equiv \pa_{i}
{\mathbb{E}^{i}}_{\mbox{\scriptsize\tt AB\normalfont\normalsize}}  +  
|[\mathbb{E}_{i}, \mathbb{E}^{i}]|_{\mbox{\scriptsize\tt AB\normalfont\normalsize}} = 0 
\label{ashgauss} 
\mbox{ } , 
\ee
\be
\mbox{tr}(\mathbb{E}^{i} \mathbb{F}_{ij}) = 0 
\label{ashmom} 
\mbox{ } , 
\ee
\be
\mbox{tr}(\mathbb{E}^{i}\mathbb{E}^{j}\mathbb{F}_{ij}) = 0 
\label{ashham} 
\mbox{ } .
\ee
(\ref{ashgauss}) arises because one is using a first-order formalism; in this particular case, it is an $SU(2)$ (Yang--Mills--)Gauss constraint.  
(\ref{ashmom}) and (\ref{ashham}) are the polynomial forms taken by the GR momentum and Hamiltonian constraints respectively, where 
$\mathbb{F}^{\mbox{\scriptsize\tt AB\normalfont\normalsize}}_{ij} := 2
\pa_{[i}\mathbb{A}^{\mbox{\scriptsize\tt AB\normalfont\normalsize}}_{j]} + 
|[\mathbb{A}_{i}, \mathbb{A}_{j}]|^{\mbox{\scriptsize\tt AB\normalfont\normalsize}}$ 
is the Yang--Mills field strength corresponding to ${\mathbb{A}_{i}}^{\mbox{\scriptsize\tt AB\normalfont\normalsize}}$. 
One can see that (\ref{ashmom}) is indeed associated with momentum since it is the condition for a vanishing (Yang--Mills--)Poynting vector.  
Again, the Hamiltonian constraint (\ref{ashham}) lacks such a clear-cut interpretation.  
Finally, the Ashtekar variables formulation given above is of complex GR.  
But then one requires troublesome \cite{Kuchar93} {\it reality conditions} in order to recover real GR.  
Nowadays so as to avoid reality conditions, one usually prefers to work not with Ashtekar's original complex variables but with Barbero's real 
variables \cite{Barbero} that now depend on an Immirzi parameter, $\gamma$. 
{\bf Loop Quantum Gravity (LQG)} is then a QM scheme for this \cite{Rovellibook, Thiemann}.
Whilst the standard constraint algebroid in this scheme is usually an enlarged but qualitatively-similar version of the Dirac algebroid, 
a {\bf Master Constraint Approach} has also been proposed \cite{Thiemann}.  
Here, the {\sl master constraint}  $\mathbb{M} := {\cal C}_{\sfA}{\cal K}^{\sfA\sfB}{\cal C}_{\sfB}/2$ (for some arbitrary positive-definite 
array ${\cal K}^{\sfA\sfB}$) is supposed to replace the set of constraints ${\cal C}_{\sfA}$.   
There are however a number of reservations with this (see each of \cite{Thiemann, APOT} and Sec \ref{FEP}), not 
least whether the arbitrariness in ${\cal K}^{\sfA\sfB}$ will cause difficulties.

\section{The Frozen Formalism Problem (FFP)} \label{FFP}

One notable facet of the POT shows up in attempting canonical quantization of GR (or of many other gravitational theories that are likewise background-independent).
It is due to GR's Hamiltonian constraint ${\cal H}$ being quadratic but not linear in the momenta.
Because of this feature, I denote the general case of such a constraint by ${\cal Q}$uad.
Then promoting a constraint with a momentum dependence of this kind to the quantum level gives a time-independent wave equation (\ref{WDE3}) that is schematically of the form 
\beq
\hat{\cal H}\Psi = 0
\label{WDE} \mbox{ } ,   
\eeq
in place of ordinary QM's time-dependent one, 
\beq
i\pa\Psi/\pa t = \hat{\fH}\Psi \mbox{ } 
\eeq 
[or a functional derivative $\delta/\delta t(\bx)$ counterpart of this in the general GR case].  
Here, $\fH$ is a Hamiltonian, $\Psi$ is the wavefunction of the universe and $t$ is absolute Newtonian time [or a a local GR-type generalization $t(\bx)$].

This suggests, in apparent contradiction with everyday experience, that nothing at all {\sl happens} in the universe.  
Thus one is faced with having to explain the origin of the notions of time in the laws of physics that appear to apply in the universe; 
this paper reviews a number of strategies for such explanations.   
Moreover timeless equations such as the WDE apply {\sl to the universe as a whole}, whereas the more ordinary laws of physics apply to small 
subsystems {\sl within} the universe.  
This suggests that the paradox apparent rather than actual.

The usual form of the FFP is that variation with respect to the lapse implies the purely quadratic ${\cal H}$, which implies the FFP at the quantum level.   
That reparametrization invariance implies a purely quadratic ${\cal H}$ is also fairly commonly stated in the literature. 
However, the stance of Barbour and collaborators \cite{BSW, RWR} is that Leibnizian Temporal Relationalism is the starting point and that this is 
{\sl implemented by} reparametrization invariance alongside an absense of auxiliary time-like variables such as Newtonian absolute time or the GR lapse.  
The subsequent product-type action, modellable in this sense by the Jacobi action of mechanics, $S = \sqrt{2}\int\d s\sqrt{E - V}$ 
(for $\d s$ the configuration space line element, $V$ the potential energy and $E$ the total energy), implies a purely quadratic constraint 
as a {\sl primary} constraint which, as ever, then gives a FFP. 
One can then view Leibnizian Temporal Relationalism as {\sl setting up} a {\bf classical FFP}, which is resolved (at least in principle, 
see Sec \ref{CR} for complications) in a `time is to be abstracted from change' Machian manner by the {\bf Jacobi--Barbour--Bertotti (JBB) emergent 
time} \cite{BB82, B94I, Kieferbook, FileR},   
\beq 
t^{\sJ\sB\sB} = t^{\sJ\sB\sB}(0) + \int\d s/\sqrt{E - V} \mbox{ } . 
\eeq
In the context of Mechanics, this arises as a simplifier of the relational approach's classical equations and amounts to a recovery of Newtonian time 
from relational first principles.
On the other hand, in the context of GR, it amounts to recovery of proper time, and of cosmic time, in the case of (approximately) homogeneous cosmologies.
However, this fails to unfreeze the quantum FFP, so this approach does not suffice by itself.  


An outline of the strategization about the FFP involves the long-standing philosophical fork mentioned in Sec 1.1 between `time is fundamental' and  
`time should be eliminated from one's conceptualization of the world'.  
A finer classification \cite{Kuchar92,I93,APOT} is into Tempus Ante Quantum, Tempus Post Quantum, Tempus Nihil Est \cite{Kuchar92, I93} and Non Tempus sed Historia \cite{APOT}. 
My separation out of the last of these from \K and Isham's timeless approaches is due to its departure from conventional physics' dynamics 
and quantization of {\sl configurations} and conjugates. 
I denote these by, procedurally ordered from left to right, {\sc tq}, {\sc qt}, {\sc q} and a new pair: {\sc hq}, {\sc qh} (`Historia ante Quantum' and `Historia post Quantum').
In fact, one has an octalemma, the final possible cases being {\sc h}, {\sc t} or none. 
There three correspond to opting out of quantizing GR, on grounds of regarding it as but an effective classical theory (much as one 
would usually not directly quantize the equations of classical Fluid Mechanics).  
In such a case, whatever `theory of true degrees of freedom' that eventually replaces GR as 
a fundamental theory, which is what one would then aim to quantize, might not possess GR-like features including Background Independence.
I refer the reader interested in such approaches to Carlip's recent review \cite{Carlip12}.
This Review concentrates on the first five above.  
One might also consider {\sc h} and {\sc t} as classical antecedents that are nevertheless interesting models in their own right 
(i.e. not abandon quantization of GR, but rather leave it instead for a future occasion).  

\mbox{ } 

\noindent Note 1) By aiming to involve a notion of time at all, {\sc tq} and {\sc qt} strategies favour ordinary QM, whereas 
Tempus Nihil Est and Non Tempus sed Historia are more radical in this respect.

\noindent Note 2) In finer detail, {\sl what} change is to be used in Mach's statement \cite{ARel, ARel2}? 
Various suggestions are 

\noindent i) {\bf any change} (Rovelli \cite{Rovellibook, Rovfqxi}, see also \cite{BojoBook, Dittrich, Dittrich2}). 

\noindent ii) {\bf All change} (Barbour \cite{Barfqxi}, following Leibniz in emphasizing the configuration of the universe as a whole, 
and holding this to be the only perfect clock).

\noindent iii) {\bf A sufficient totality of local relevant change (STLRC)}, which is my position, in which all changes 
are given the opportunity to contribute but only those that do so locally significantly are kept in the actual time calculation, 
as exemplified by the actual calculations of the astronomers' {\it ephemeris time}.  
See \cite{ARel, ARel2, ACos2} for further detail of these, as well as further detail of how to classify POT strategies by this criterion.  
The above $t^{\sJ\sB\sB}$ {\sl looks} to be of the `all' type, but, upon detailed examination \cite{FileR, ARel2}, it is in practise of the STLRC form. 

\noindent Note 3) In these various approaches to timekeeping, one should be careful not to confuse a convenient reading-hand 
(this is usually small: a conventional clock-like device) with what system is actually used to maintain calibration of one's 
timestandards/clocks (this is often much larger, such as the Earth--Moon--Sun system).    

\noindent Note 4) One might also expect it only to make sense for one subsystem to furnish a time for another if the two are coupled dynamically 
or by time-frame determination procedure, else how would the one subsystem know about the other one it is keeping time {\sl for}.  
Some programs make no use of coupling between their clock subsystem and their subsystem under study \cite{PW83, Rovellibook}.  
However, as we shall see in Sec 6.2, other programs do require a nonzero coupling; conceptually, this corresponds to the subsystem under study 
needing to be aware of the physics of the clock subsystem in such programs.

\section{The Problem of Time Possesses Other Facets} \label{Other}

\noindent Over the past decade, it has become more common to suggest or imply that the POT {\sl is} the Frozen Formalism Problem.  
However, a more long-standing point of view \cite{Kuchar91, Kuchar92, I93, Kuchar93} (and also argued in favour for in e.g. 
\cite{Kuchar99, Kieferbook, APOT, FileR} and the present article) is that the POT contains a number of further facets.  
Furthermore, it is not even a case of then having to address around eight facets in succession.
For, as Kucha\v{r} found \cite{Kuchar93}, these facets interfere with each other.  
I argue \cite{APOT, FileR} that this occurs because the facets arise from a common cause -- the mismatch of the notions of time in GR and Quantum Theory -- 
by which they bear conceptual and technical relations, making it advantageous, and likely necessary for genuine progress, to treat them as a coherent 
package rather than piecemeal as `unrelated problems'.

I also note that, despite the facets' common origin, almost all of the strategizing toward overcoming the POT has started from how to address the FFP. 
This is largely how POT approaches have been classified and conceptualized about. 
Whilst it does make for particularly eye-catching conceptualization, I would argue, rather, that 

\noindent 1) each facet should directly receive conceptual and strategic consideration, for which this Review is a first step.  

\noindent 2) That it is likely that only some {\sl combined} strategy will stand much of a chance of overcoming most of, or even all of, the POT; 
this Review's last Section considers a few combined strategies.

\noindent There are then seven further facets to examine, alongside various of the inter-relations between them.  
The FFP itself has the following addendum.  

\mbox{ }  

\noindent 
{\bf Hilbert Space/Inner Product Problem}, i.e. how to turn the space of solutions of the frozen equation in question into a Hilbert space.
See Sec \ref{TPQ} for why this is a problem (i.e. why ordinary QM inspired guesses for this will not do for GR).
It is a time problem due to the ties between inner products, conservation and unitary evolution.

\section{Further facets of the Problem of Time}\label{Facets}

\noindent {\bf Best Matching Problem} \cite{BB82, RWR}.  
If one's theory exhibits Configurational Relationalism, linear constraints ${\cal L}\mi\mn_{\sfZ}$ follow from variation with respect to the 
auxiliary $G$-variables [these constraints generalize (\ref{Mom}) in the $G$ = Diff($\Sigma$) case of GR].  
The Best Matching procedure is then whether one can solve the Lagrangian form of 
${\cal L}\mi\mn_{\sfZ}\lfloor \ttQ^{\sfA}, \ttP_{\sfA}\rfloor = 0$ 
for the $G$-auxiliaries themselves; this is a particular form of reduction.
One is to then substitute the answer back into the action. 
This is clearly a classical-level procedure, an indirect implementation of Configurational Relationalism that is a bringing into  
maximum congruence by keeping $\ttQ^{\sfA}_1$ fixed and shuffling $\ttQ^{\sfA}_2$ around with $G$-transformations until it is as close to $\ttQ^{\sfA}_1$ as possible.
From another perspective, one is establishing a point identification map \cite{Stewart} between configurations.  
It becomes a Problem due to being an often-obstructed calculation.  
In particular this is so in the case of GR (or, more widely, theories with a Diff-type $G$) where it has been more usually called the 
{\bf Sandwich Problem} \cite{BSW, W63, BO69, BF93, FodorTh}, in reference to the construction of the spacetime `filling' between 2 given spatial hypersurfaces (`slices of bread').  
\noindent See Sec \ref{CR} for further generalization of this facet.
\noindent One relation of this to the POT is clear via the title of \cite{BSW} is ``{\it three-dimensional geometry as carrier of information about time}"; 
furthermore, it becomes entwined in some POT approaches that start at the pre-quantum level (see Sec \ref{CR}).

\mbox{ } 

\noindent {\bf Classically, there are no Closure, Foliation-Dependence or Spacetime Reconstruction Problems.}    
These are all 

\noindent ensured by the nature of the Dirac algebroid. 
From (\ref{mommom}-\ref{hamham}), closure is clear.
In classical GR, one can foliate spacetime in many ways, each corresponding to a different choice of timefunction.  
This is how time in classical GR comes to be `many-fingered', 

\noindent with each finger `pointing orthogonally' to each possible foliation.  
Classical GR then has the remarkable property of being refoliation-invariant \cite{HKT}, so that going between two given spatial geometries by means 
of different foliations in between produces the same region of spacetime and so the same answers to whatever physical questions can be posed therein.
The `relativity without relativity' (RWR) approach \cite{RWR} then amounts to a classical-level spacetime reconstruction: here spacetime structure 
is not presupposed, only Configurational Relationalism and Temporal Relationalism, and then the Dirac procedure for constraint algebroid consistency 
returns GR spacetime as one of very few consistent possibilities. 
[GR is further picked out among these by the demand of foliation independence and the demand for a finite nonzero propagation speed.]  
RWR includes how, for {\sl simple} \cite{Phan, Lan2} matter, local SR is deduced rather than assumed \cite{RWR}.
Moreover, following on from Sec 1, RWR shows how GR can not only be cast in Machian form, but also can be derived from Machian principles 
that assume less structure (though consult \cite{Phan, Lan2} for an update of what assumptions this approach actually entails).
Finally, RWR is an answer to Wheeler's question about why ${\cal H}$ takes the form that it does (see \cite{HKT} for an earlier 
answer, though that did assume embeddability into spacetime and thus does not constitute a classical-level spacetime reconstruction).  
However, one then no longer knows any means of guaranteeing these three nice properties at the quantum level.  

\mbox{ } 

\noindent {\bf QM Foliation Dependence Problem}. 
That this is obviously a time problem follows from each foliation by spacelike hypersurfaces being orthogonal to a GR timefunction: 
each slice can be interpreted as an instant of time for a cloud of observers distributed over the slice, and each foliation 
corresponds to the these moving in a particular way.  
\noindent As \K states \cite{Kuchar92}, {\it ``When one starts with the same initial state} $\Psi_{\si\sn}$ {\it on the initial hypersurface 
and develops it to the final hypersurface along two different routes} $\Psi_{\sf\si\sn-1} \neq \Psi_{\sf\si\sn-2}$ {\it on the final hypersurface.  
Such a situation certainly violates what one would expect of a relativistic theory."} 
A further issue here is how observers operationally determine the foliation for the physics that they experience.   

\mbox{ }

\noindent {\bf QM Constraint Closure Problem = Functional Evolution Problem}
Closure of a classical Poisson bracket algebroid does not entail closure of the corresponding quantum commutator algebroid, due to 
1) these algebroids not necessarily being isomorphic due to global effects \cite{I84}. 
2) the existence of anomalies (by definition quantum obstructions arising in cases that did not possess classical obstructions); 
Dirac \cite{Dirac} said that avoiding this required luck. 
The Functional Evolution Problem is then a subset of the possibility of anomalies due to time/frame issues in GR or some toy model of or alternative thereto. 
[It is a subset since only some of the anomalies that one finds in physics are time- or frame-related.]    
Foliation-dependent anomalies clearly exhibit {\sl two} Facets of the POT.  
\noindent [Non-closure is also entwined with the Operator Ordering Problem, since changing the ordering gives additional right-hand-side 
pieces not present in the classical Poisson brackets constraints algebroid.]  

\mbox{ } 

\noindent  {\bf Global Problem of Time} alias {\bf Kucha\v{r}'s Embarrassment of Poverty} \cite{Kuchar92}.  
This is a more basic problem, which is already present at the classical level (recall Sec 1), as an obstruction to making 
globally-valid gauge choices in parallel with the well-known Gribov effect of Yang--Mills theory, except that in the 
present situation, gauge choices involve choices of times and frames.
Its subfacets are, in one sense,
\noindent i) timefunctions may not be globally defined in space.
\noindent ii) Timefunctions may not be globally defined in time itself.  
\noindent In another sense, at the classical level this might be seen as a need to use multiple coordinate patches held together by the meshing 
conditions of classical differential geometry.
Whilst that is classically straightforward, it remains largely unclear how one might formulate a `meshing together' of quantum evolutions themselves.  

\mbox{ } 

\noindent {\bf Multiple Choice Problem} alias {\bf Kucha\v{r}'s Embarrassment of Riches} \cite{Kuchar92}.  
This is the purely quantum-mechanical problem that different choices of time variable may give inequivalent quantum theories.  
\noindent There is a notion of time is among the coordinates in GR; in some approaches, it is among the observables.
\noindent Foliation Dependence is one of the ways in which the Multiple Choice Problem can manifest itself.
Moreover, the Multiple Choice Problem is known to occur even in some finite toy models \cite{Kuchar92}. 
Thus foliation issues are not the only source of the Multiple Choice Problem.  
For instance, another way the Multiple Choice Problem can manifest itself is as a subcase of how making different choices of sets 
of variables to quantize (as per the Groenewold--Van Hove phenomenon \cite{Gotay}) may give inequivalent quantum theories.    

\mbox{ } 

\noindent  {\bf Problem of Beables}. 
This is usually called the Problem of Observables \cite{Dirac, Kuchar92, I93, Rovellibook}, but I follow Bell \cite{Speakable, Bell, Speakable2} in 
placing emphasis on conceiving in terms of beables, which carry no connotations of external observing, but rather simply of being, 
which make them more suitable for the quantum-cosmological setting.  
The Problem involves construction of a sufficient set of beables for the physics of one's model, which are then involved in the model's notion of evolution.  

\mbox{ } 

\noindent {\bf Spacetime (Reconstruction or Replacement) Problem}.  
\noindent 1) At the at the quantum level, fluctuations of the dynamical 

\noindent entities are unavoidable, i.e. here fluctuations of 3-geometry, 
and these are then too numerous to be embedded within a single spacetime (see e.g. \cite{Battelle}).  
Thus (something like) the superspace picture -- considering the set of possible 3-geometries -- might be expected to take over from the spacetime picture at the quantum level.  
It is then not clear what becomes of causality (or of locality, if one believes that the quantum replacement for spacetime is `foamy' \cite{Battelle}); 
in particular, microcausality is violated in some such approaches \cite{Savvidou04}.  

\noindent 2) {\bf Recovering continuity and}, a forteriori, {\bf something that looks like spacetime} (e.g. as regards dimensionality) is 
\noindent an issue in discrete or bottom-up approaches to Quantum Gravity. 
This is not a given, since some approaches give unclassical entities or too low a continuum dimension.  
See Sec \ref{SRP} for a third subfacet.

\section{More detail of a selection of Frozen Formalism Problem Strategies}\label{FFP-Strat}

{            \begin{figure}[ht]\centering\includegraphics[width=1.0\textwidth]{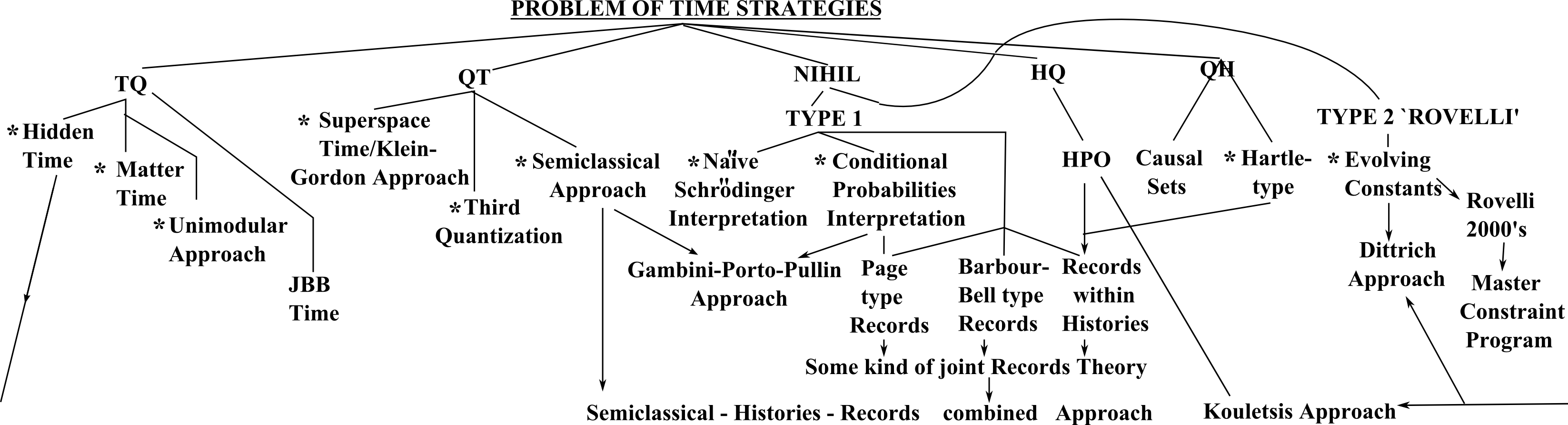}
\caption[Text der im Bilderverzeichnis auftaucht]{        \footnotesize{The various branches of 
strategy and their relations, in a diagram with `cylindrical topology'. 
* indicates the 10 strategies covered by Kucha\v{r} and Isham's classic reviews \cite{Kuchar92, I93}.}  } \label{Fig43}\end{figure}          }
%
Fig 1 further refines Sec \ref{FFP}'s classification.
This includes a dichotomy of Tempus Nihil Est between Rovelli's school \cite{Rov91, Rov02, Rovellibook, Dittrich} and the somewhat broader and 
partly older schools of Hawking, Page, Wootters, Barbour, Gell-Mann, Hartle, Halliwell and I \cite{HP86, PW83, Page1, Page2, GMH, B94II, 
EOT, H99, H03, Records, H09, AHall}.  
This Review mostly only covers the latter (though see Sec \ref{Beables} and \cite{APOT} for comments about the former).  
The present Sec very largely ignores Configurational Relationalism/the ensuing linear constraints, which situation is rectified in the subsequent Sec.
Thus suitable toy models for the present section include {\bf minisuperspace} (GR restricted to the homogeneous solutions, 
which also entails no possibility of structure formation) and Sec 3's relational JBB formulation of mechanics.

\subsection{Tempus Ante Quantum ({\sc tq})}\label{TAQ}

{\bf Ante Postulate}.  There is a fundamental time to be found at the classical level for the full (i.e. untruncated) classical gravitational theory 
(possibly coupled to suitable matter fields).  
Candidate times of this type include the following.

\mbox{ }

\noindent 1) Whilst $\tea^{\sJ\sB\sB}$ is obtained in accordance with the Ante Postulate, it is not an unfreezing at the quantum level.  
Thus it is only a classical resolution of the FFP.

\mbox{ }  

\noindent 2) {\bf Riem time}.\footnote{This is usually called superspace time, though it is only that for minisuperspace, for which Riem and superspace coincide}
Perhaps one could then take the indefinite direction to be pick out a timefunction, in parallel to how the indefinite direction in Minkowski space does for such as Klein--Gordon theory. 
Schematically, 
\beq
\triangle_{\mbox{\scriptsize \boldmath${\cal G}$}} 
\mbox{ }  \mbox{ } \mbox{ is actually a } \mbox{ }   \Box_{\mbox{\scriptsize \boldmath${\cal G}$}} \mbox{ } .  
\eeq
This approach was traditionally billed as Tempus Post Quantum, through choosing to make this identification of a time after the quantization, 
but the approach is essentially unchanged if this identification is made priorly at the classical level. 
Regardless of when this identification is made, however, this approach fails by \cite{Kuchar81, Kuchar91} 
the GR potential not being as complicit as Klein--Gordon theory's simple mass term was (it does not respect the GR configuration space's conformal Killing vector).

\mbox{ }  

\noindent 3) The above is one example of a scale time.  
\noindent Scale or related quantities are often used as internal times, for instance the cosmological scalefactor $a$, or the local spatial 
volume element $\sqrt{h}$ that goes as $a^3$ in the (approximately) isotropic case. 

\mbox{ } 

\noindent {\bf Part-linear/parabolic implementation}.  
While the above straightforward schemes fail, it may still be that a classical time exists but happens to be harder to find.
We now consider starting one's scheme off by finding a way of solving ${\cal Q}\mbox{uad}$ to obtain a part-linear/parabolic form, 
where I make a single-time versus time-and-frame formulation distinction 
\beq
\ttP_{\stea^{\ta\tn\ttt\te}} + 
H^{\st\sr\su\se}\lfloor \tea^{\sa\sn\st\se}, \ttQ_{\so\st\sh\se\sr}^{\sfA}, \ttP^{\so\st\sh\se\sr}_{\sfA}\rfloor  = 0 
\mbox{ } \mbox{ or } \mbox{ } \ttP^{{\cal X}}_{\mu} + 
H^{\st\sr\su\se}_{\mu}\lfloor {\cal X}^{\mu}, \ttQ_{\so\st\sh\se\sr}^{\sfA}, \ttP^{\so\st\sh\se\sr}_{\sfA}\rfloor = 0 \mbox{ } .
\label{PLIN}
\eeq
Here, $\ttP_{\stea^{\ta\tn\ttt\te}}$ is the momentum conjugate to a candidate classical time variable, $\tea^{\sa\sn\st\se}$ which 
is to play a role parallel to that of external classical time.  
Also, $\ttP_{\cal X}^{\mu}$ are the momenta conjugate to 4 candidate embedding variables ${\cal X}^{\mu}$ (a 4-vector time-and-frame quantity  
$[\tea^{\st\sr\su\se}, \mX^{i}]$, where $\mX^{i}$ are 3 spatial frame variables). 
$H^{\st\sr\su\se}$ is then the `true Hamiltonian' for the system.  
Also, $H^{\st\sr\su\se}_{\mu} = [H^{\st\sr\su\se}, \Pi^{\st\sr\su\se}_{i}]$, where $\Pi_{i}^{\st\sr\su\se}$ is the true momentum flux 
constraint.
By passing as soon as possible to having an object that plays such a role, {\sc tq} can be viewed as the most conservative family of strategies \cite{I93, T97}.
Given such a parabolic form for ${\cal H}$, it is then possible to apply a conceptually-standard quantization that yields the time-dependent Schr\"{o}dinger equation,  
\beq
i\ttD\Psi/\ttD \tea^{\sa\sn\st\se} = \hat{H}_{\st\sr\su\se}\lfloor \tea^{\sa\sn\st\se}, \ttQ_{\so\st\sh\se\sr}^{\sfA}, \ttP^{\so\st\sh\se\sr}_{\sfA}\rfloor\Psi 
\mbox{ } \mbox{ or } \mbox{ }    
i\ttD\Psi/\ttD {\cal X}^{\mu} = \hat{H}^{\st\sr\su\se}_{\mu}
\lfloor {\cal X}^{\mu}, \ttQ_{\so\st\sh\se\sr}^{\sfA}, \ttP^{\so\st\sh\se\sr}_{\sfA}\rfloor\Psi \mbox{ } .   
\label{TEEDEE}
\eeq 
Here, $\ttD$ is the portmanteau of partial derivative $\pa$ for finite theories and functional derivative $\delta$ for field theories.
The above unfreezing is accompanied, at least formally, by the obvious associated Schr\"{o}dinger inner product.  
Single-time and time-and-frame part-linear forms then occur for parametrized nonrelativistic particle and parametrized field theory toy models respectively \cite{Kuchar92}.  
GR examples of part-linear forms are as follows.

\mbox{ }  

\noindent 4) {\bf Hidden time Internal Schr\"odinger Approach}
A first such suggestion is that there may be a {\it hidden} alias {\it internal} time \cite{York72, Kuchar81, Kuchar92, I93, GT11} within one's gravitational theory itself.  
I.e. that the apparent frozenness is a formalism-dependent statement, to be removed via applying some canonical transformation.
In the first part-linear form scheme, this sends GR's spatial 3-geometry configurations to 1 hidden time, so $\mt^{\sa\sn\st\se} = \mt^{\sh\si\sd\sd\se\sn}$ 
plus 2 `true gravitational degrees of freedom' [which are the form the `other variables' take, and are here `physical' alias `non-gauge'].
The general canonical transformation is here 
\beq
(\ttQ^{\sfA}, \ttP_{\sfA}) \longrightarrow 
(\tea^{\sh\si\sd\sd\se\sn}, \ttp_{\stea^{\th\ti\td\td\te\tn}}, \ttQ_{\st\sr\su\se}^{{\sfA}}, \ttP^{\st\sr\su\se}_{{\sfA}}) 
\mbox{ } \mbox{ } \mbox{or} \mbox{ } \mbox{ }
(h_{ij}(x^k), \pi^{ij}(x^k))  \longrightarrow 
(\chi^{\mu}(x^{k}), \Pi_{\mu}(x^{k}), \mQ_{\mbox{\scriptsize true}}^{\sfA}(x^{k}), \mP^{\mbox{\scriptsize true}}_{\sfA}(x^{k})) 
\mbox{ } , 
\label{canex} 
\eeq
for $\mQ_{\mbox{\scriptsize true}}^{\sfA}(x^{i})$ the true gravitational degrees of freedom and embedding variables $\chi^{\mu}$.
Thus one arrives at a hidden time-dependent Schr\"{o}dinger equation of whichever of the two forms in (\ref{TEEDEE}), 
with the above significances attached to the variables (for $\tea^{\sh\si\sd\sd\se\sn} = \tea^{\sa\sn\st\se}$ or $\chi^{\mu} = {\cal X}^{\mu}$).

A particular sub-example of this involves the canonical conjugate of $\sqrt{h}$, namely the {\it York time} (\cite{York72}). 
This is proportional to the constant mean curvature (CMC) ($K$ = constant and thus $h_{ij}\pi^{ij}/\sqrt{h}$ = constant$^{\prime}$, and only applies in solutions that 
admit foliations by CMC slices.  
In this shape-scale split formulation, the GR constraints decouple on CMC slices so that one can at least formally solve the momentum constraint prior to the Hamiltonian constraint.
This approach's furtherly-reduced configuration space is the space of conformal 3-geometries: conformal superspace CS($\Sigma$) := Superspace($\Sigma$)/Conf($\Sigma$), 
to which a solitary global spatial volume of the universe variable is usually adjoined, for Conf($\Sigma$) the conformal transformations on $\Sigma$.  
This approach is in practice hampered by Example 1.2 of Sec \ref{Global} and by the Lichnerowicz--York equation (conformalized Hamiltonian constraint) not being explicitly solvable.
Substantial progress with shape--scale split formulations in the case of Ashtekar variables has been but recent \cite{BST12}.

\mbox{ }  

\noindent 5) {\bf Straightforward Matter Time}.  
Typically in minisuperspace for 1-component scalar matter, one simply isolates the corresponding momentum to play the role of the time part of the subsequent wave equation.  
This often taken to be `the alternative' to scale time [e.g. this is often so in the {\bf Loop Quantum Cosmology (LQC)} literature, which is a minisuperspace analogue but 
now based on a higher-order difference equation form of WDE].
However, the momenta conjugate to each of these represent 2 further possibilities, and then, if if canonical transformations are allowed, a whole further host of 
possibilities become apparent.  
Finally, using a matter scalar field as a time is often argued to be `relational' \cite{Bojowald05}, though this is clearly only the `any change' form of the 
Machian recovery of time, so more scrutiny might be considered. 
[For each such model, does this candidate matter time possess the features expected of a time?  
If there is multi-component matter, why should one matter species be given the privilege of constituting the time? 
If one uses other times -- {\sl equally} relational in the above sense -- does the Multiple Choice Problem appear?]

\mbox{ } 

\noindent 6) {\bf Reference Matter Time}. a) One can look to attain the part-linear form i) of (\ref{canex}) is attained with $\mt^{\sa\sn\st\se} = 
\mt^{\sm\sa\st\st\se\sr}$ and $\mQ^{\Gamma}_{\so\st\sh\se\sr} =$ $h_{ij}$, or (better) the 4-component version, by extending the geometrodynamical 
set of variables to include matter variables coupled to these, which then serve so as to label spacetime events \cite{Kuchar92}.
b) One could additionally form a quadratic combination of constraints \cite{BK94KR95, BroMa}, resulting in strongly-vanishing Poisson brackets. 
Technical problems with these include involving a choice between time resolution and physicality of the matter, and the more general non-correspondence 
between reference matter and tangible, observed matter; see e.g. \cite{Kuchar92, APOT, Book} for further details.  
\cite{HP11} exemplifies newer such approaches using both further types/formulations of matter and the Ashtekar variables formulation of GR.

\mbox{ } 

\noindent 7) {\bf Unimodular gravity implementation}
\noindent One might also regard an undetermined cosmological constant $\Lambda$ as a type of reference fluid \cite{UW89}.
Here, one does not consider the lapse to be a variable that is to be varied with respect to.  
Instead, ${\cal H}_{i}$ has ${\cal H}_{,i}$ as an integrability [c.f. (\ref{hamham})], leading to ${\cal H} + 2\Lambda = 0$ with 
${\cal H}$ the vacuum expression and $\Lambda$ now interpreted as a constant of integration. 
The unimodular approach has problems of its own as a POT resolution (one is in Sec \ref{Fol}; see \cite{Kuchar92, I93, FileR} for further such).

\mbox{ } 

\noindent Aside from the individual technical problems of each of the above approaches, finding or appending a time is contrary to inherent primary Leibniz 
timelessness and subsequent Machian time emergence.  
Furtherly, these times are particular special changes rather than allowing all changes to contribute in principle as per the STLRC interpretation of Machian time emergence.  
E.g. matter change has no opportunity to contribute to scale, York or unimodular times, whilst gravitational change has no opportunity to contribute to matter time.

\mbox{ } 
 
\noindent {\bf Machian Status Quo}.  This is based on {\sl current knowledge} of {\sc t} ... {\sc q} approaches that remain unfreezing at the 
quantum level and how these are all characterized as un-Machian and bereft of accuracy.    
Thus {\sc q} ... {\sc t} {\sl looks to be presently required for a satisfactory QM-level Machian time} in the STLRC sense.

\subsection{Tempus Post Quantum ({\sc qt})}\label{TPQ}

\noindent {\bf Post Postulate} In strategies in which time is not always present at the fundamental level, time is nevertheless capable of emerging in the quantum regime.  
Because this is an emergence, it means that the Hilbert space structure of the final quantum theory is capable of being (largely) unrelated to that of the 
WDE-type quantum theory that one starts with.   
Such emergent strategies are of the following types. 

\mbox{ }

\noindent 1) A scheme based on Schr\"{o}dinger inner product fails due to the indefiniteness of the WDE.

\noindent 2) {\bf Attempting a Klein-Gordon Interpretation based on Riem time} then fails just as for its {\sc tq}-ordered case.
However, noting the parallel with Klein--Gordon theory failing as a first-quantization leads to its reinterpretation as a second-quantized QFT, leads one to try the following 
for the WDE.  

   
\noindent 3) {\bf Third Quantization}, i.e. that the solutions $\Psi[h]$ of the WDE might be turned into operators:
\beq
\widehat{\cal H}\widehat{\Psi}\psi = 0 \mbox{ } .  
\eeq
However, this turns out not to shed much light on the POT \cite{Kuchar92}, and, whilst Third Quantization recurs \cite{GFT} in 
the Group Field Theory approach to Spin Foams (themselves described in Sec \ref{Histor}), its use there is not extended to furnish a POT strategy either.

\mbox{ }          

\noindent 4) {\bf Semiclassical approach} \cite{DeWitt67, HallHaw, Kuchar92, I93, Kieferbook, SemiclI, SemiclIII, ACos2}.
Perhaps one has slow, heavy `h'  variables that provide an approximate timestandard with respect to which the other fast, light `l' 
degrees of freedom evolve \cite{HallHaw, Kuchar92, Kieferbook}.  
In the Halliwell--Hawking \cite{HallHaw} scheme for GR Quantum Cosmology, h is scale (and homogeneous matter modes) and the l-part are 
small inhomogeneities.\footnote{This is a quantum explanation for the origin of structure in the universe 
- the seeding of galaxies and of CMB inhomogeneities.    
The possibility of making a connection between quantum-cosmological perturbations and the observed universe is usually via some inflationary 
mechanism, and renders this semiclassical Quantum Cosmological regime particularly valuable among quantum-gravitational 
regimes from the perspective of future observations.  
This example only makes sense once linear constraints are included, i.e. it is a `perturbations about minisuperspace' midisuperspace treatment. 
This study illustrates that whilst evoking semiclassicality is in some ways a limitation, semiclassical approaches nevertheless have valuable content. \label{gnimmel}}
%
The most usual Semiclassical Approach for a POT strategy involves making i) the Born--Oppenheimer ansatz 
$\Psi(\mh, \ml) = \psi(\mh)|\chi(\mh, \ml)\rangle$,
\beq
\mbox{ii) the WKB ansatz} \hspace{2in}
\psi(\mh) = \mbox{exp}(i\, S(\mh)) \mbox{ } ; \hspace{5in}  
\label{WKB}
\eeq 
in each case one makes a number of associated approximations.

\noindent iii) One forms the h-equation $\langle\chi| \hat{\cal H} \Psi = 0$.  
Then, under a number of simplifications, this yields a Hamilton--Jacobi equation $\{\pa S/\pa\mh\}^2 = 2\{E - V(\mh)\}$, where $V(\mh)$ is the h-part of the potential. 
One way of solving this involves doing so for an approximate emergent semiclassical time $t^{\se\sm(\sW\sK\sB)}(\mh)$; this is a recovery of a 
`time before quantization' timefunction, namely $\tea^{\sJ\sB\sB}$; moreover, this now does have a QM-unfreezing status in this new emergence.  

\noindent iv) One then forms the l-equation $\{1 - |\chi\rangle\langle\chi|\}\hat{\cal H}\Psi = 0$. 
This fluctuation equation can be recast (modulo further approximations) into a $\tea^{\sW\sK\sB}$-dependent Schr\"{o}dinger equation for the l-degrees of freedom,
\beq
i\ttD|\chi\rangle/\ttD \tea^{\sW\sK\sB}  = \widehat{\cal H}_{\sll}|\chi\rangle \mbox{ }
\label{TDSE2}   \mbox{ } . 
\eeq
The emergent time dependent left-hand side of this arises from the `chroniferous' (time bearing) cross-term $\pa_{\sh}|\chi\rangle\pa_{\sh} \psi$  
(e.g. for a mechanics model with a single h with conjugate momentum p$_{\sh}$ and configuration space metric $\bM$ with inverse $\bN$), by the following combination of basic moves.  
\beq
N^{\sh\sh}  i  \frac{\pa\fW}{\pa \mh}\frac{\pa \left| \chi\right \rangle}{\pa \mh} = 
i N^{\sh\sh} \mp_{\sh}  \frac{\pa \left| \chi\right \rangle}{\pa \mh} =
i N^{\sh\sh}  M_{\sh\sh}\frac{\d \mh}{\d t^{\sW\sK\sB}}      \frac{\pa \left| \chi\right \rangle}{\pa \mh} = 
i \frac{  \pa \mh       }{  \pa t^{\sW\sK\sB}  }\frac{\pa \left| \chi\right \rangle}{\pa \mh} = 
i \frac{\pa \left| \chi\right\rangle }{  \pa t^{\sW\sK\sB} }   \mbox{ } .  
\label{chroni}
\eeq
To finish interpreting (\ref{TDSE2}), $\widehat{\cal H}_{\sll}$ is the remaining surviving piece of $\widehat{\cal H}$, acting as a Hamiltonian for the l-subsystem.  

\mbox{ } 

\noindent Note 1) The working leading to such a time-dependent wave equation ceases to function in the absence of making the WKB ansatz and approximation. 

\noindent Note 2) {\bf WKB Problem}.  Moreover, making the WKB approximation in the quantum-cosmological context requires justification 
(see e.g. \cite{SemiclI, FileR} including for references to the original literature).
Without this justification there is a serious danger of `merely passing the buck' from a time to a set of wavefronts very heavily implying a time.  

\noindent Note 3) In the Semiclassical Approach, the time-provider to studied-subsystem coupling is obligatory ($\chi$ must be a nontrivial function of h as well as l), 
since if not, the emergent time derivative in eq (\ref{TDSE2}) would be built from a product containing a zero factor in it [the second derivative factor in each term of (\ref{chroni})].

\noindent Note 4) The first approximation used here is rather un-Machian (in the STLRC sense) via deriving its change just from scale 
(plus possibly homogeneous isotropic matter modes, or, more widely, from the usually-small subset of h degrees of freedom). 
However, the second approximation \cite{ACos2} remedies this by allowing anisotropic and inhomogeneous changes (or, more widely, 
whatever else the l-degrees of freedom may be) to contribute to the corrected emergent time.  
Contrast this with the criticism at the end of Sec \ref{TAQ}.  
Finally, the detailed emergent time here is {\sl not} exactly the same as the detailed classical $\tea^{\sJ\sB\sB}$, as is to be 
expected from how {\sl quantum} change corrections influence this new quantum-level version.  

\noindent Note 5) The qualitative types of often-omitted terms in semiclassical Quantum Cosmology are \cite{MP98, SemiclI, SemiclIII, Arce, ACos2}: non-adiabaticities, 
other (including higher) emergent time derivatives and averaged terms .  
Including the last of these parallels the use of Hartree--Fock self-consistent iterative schemes, though the system is now more complex 
via involving a chroniferous quantum-average-corrected Hamilton--Jacobi equation.  

\noindent Note 6) The imprecision due to omitted terms entails deviation from exact unitarity.

\subsection{Tempus Nihil Est}\label{TNE}

\noindent {\bf Temporal and Atemporal Questions}.  
\noindent Questions are closely related to {\sl propositional logic} (see \cite{Mackey, IL2, IS02, FileR, ARel} for applications of this to fundamental physics).  

\noindent{\bf Questions of Being} are covered by some atemporal logic.  
Among these, {\bf Questions of Conditioned Being} play a particular role.  
These involve two properties within a single instant: given that an (approximate) (sub)configuration has property $P_1$, what is the probability that it also has property $P_2$?  

\noindent Moreover, in the presence of a meaningful notion of time, one can additionally consider the following question-types to be primary.

\noindent I) {\bf Questions of Being at a Particular Time} have the general form Prob($S_1$ has property $P_1$ as the timefunction $\tea$ takes a fixed value $\tea_1$).

\noindent II) {\bf Questions of Becoming}, on the other hand, furthermore involve a given particular (approximate) 
(sub)system state becoming some other such.
As such, they are covered by a more complicated temporal logic.  

\noindent Moreover, 1) is straightforward to eliminate and some schemes for eliminating 2) too have also been put forward. 
If these things are possible, one would then expect \cite{Records} the form of the remaining physics to strongly reflect the structure of atemporal logic.
(At the classical level, Boolean logic suffices, but this is not obeyed by quantum propositions in general, leading to suggestions for the use of 
`quantum logic' \cite{I94, IL2} or the `intuitionistic logic' of Topos Theory \cite{ID, I10}.)  

\mbox{ }

\noindent{\bf Nihil Postulate}. One aims to supplant `becoming' with `being' at the primary level \cite{Page1, Page2, B94II, EOT, GMH, H99, 
HT02, H03, Records, H09}.   
In this sense, Timeless Approaches consider the instant/space as primary and spacetime/dynamics/history as secondary.

\noindent Adopting a Tempus Nihil Est approach allows one to avoid the difficult issue of trying to define time as outlined in Sec 1.  
However, one then has to face three other problems instead.

\mbox{ }

\noindent 1) {\bf Semblance of Dynamics Problem} \cite{B94II, EOT, H03, Kieferbook}. How to explain the semblance of dynamics if the universe is timeless as a whole.
Dynamics or history are perhaps now to be {\sl apparent notions} to be constructed from one's instant, though possibly also 
Histories Theory could provide an explanation (at the cost of more structure being assumed), see Sec 6.4.
\noindent Fig 2 gives a split into Type 1 and Type 2 `Rovelli' Tempus Nihil Est approaches.   
Type 1 includes various forms of timeless Records Theory have appeared, including Barbour's \cite{B94II}, Page's \cite{Page1, Page2}, 
mine \cite{Records, FileR, ARel2}, and also Records Theory within Histories Theory (work of Halliwell \cite{H99} building on 
work of Gell-Mann and Hartle \cite{GMH}); see below for more. 
Type 2 `Rovelli' is briefly covered in Sec 8; this is based on Rovelli's `any change' relational approach to time.     


\noindent 2) {\bf Nonstandard Interpretation of Quantum Theory}.  
Type 1 Tempus Nihil Est and Histories Theory both come to involve in general questions about the interpretation of Quantum Theory, 
in particular as regards whole-universe replacements for standard Quantum Theory's Copenhagen Interpretation.
I note that some criticisms of Timeless Approaches \cite{Kuchar92, Kuchar99} are subject to wishing to preserve aspects of the 
Copenhagen Interpretation, which may not be appropriate for Quantum Cosmology and Quantum Gravity.


\noindent 3) {\bf WDE Dilemma}. Such approaches either1) invoke the WDE and so inherit some of its problems, or they do not, 
thus risking the alternative problem 2) of being incompatible with it, so that the action of the WDE operator kicks purported solutions out of the physical solution space \cite{Kuchar92}. 

\mbox{ } 

\noindent {\bf Na\"{\i}ve Schr\"{o}dinger Interpretation (NSI)} \cite{HP86, UW89}.  
This first example of a Nihil strategy concerns the `being' probabilities for universe properties such as `what is the probability that the universe is large?' 
One obtains these via consideration of the probability that the universe belongs to region $\mR$ of the configuration space that corresponds to 
a quantification of a particular such property, 

\noindent
\beq
\mbox{Prob}(\mR) \propto \int_{\sR}|\Psi|^2{\cal D}\ttQ \mbox{ } , 
\eeq 
for ${\cal D} \ttQ$ the configuration space volume element.
This approach is termed `na\"{\i}ve' due to it not using any further features of the constraint equations; 
it also involves generally non-normalizable probabilities, which, however, support finite {\sl ratios} of probabilities.  
Its implementation of propositions is Boolean via how the classical regions enter the integrals; this is problematic as per this Subsection's first paragraph.

\mbox{ } 

\noindent{\bf Proposition--Projector Association}.
The aim here is to represent propositions at the quantum level by projectors (taken to include beyond \cite{IL2} the usual context and 
interpretation that these are ascribed in ordinary Quantum Theory). 
This is in contradistinction to the above kind of attempts at representation via regions of integration.  

\mbox{ } 

\noindent In ordinary QM, for state density matrix $\Rho$ and proposition $P$ implemented by projector $\widehat{\mP}$, Prob($P$; $\Rho$) 
= tr($\widehat{\Rho}\widehat{\mP}$) with Gleason's theorem providing strong uniqueness criteria for this choice of object from the perspective of satisfying the 

\noindent basic axioms of probabilities (see e.g. \cite{Ibook}).
The formula for conditional probability in ordinary Quantum Theory is then \cite{Ibook}
\beq
\mbox{Prob}(B \in b \mbox{ at } \tea = \tea_2 | A\in a \mbox{ at } \tea = \tea_1;   \Rho) = 
\frac{     \mbox{Tr}\big(\mP^B_b(\tea_2)\,\mP^A_a(\tea_1)\,\Rho\,\mP^A_a(\tea_1)\big)         }
     {     \mbox{Tr}\big(\mP^A_a(\tea_1)\,\Rho\big)                                     } \mbox{ } .         
\label{Pr:std}
\eeq
Here, $\mP^A_a$ is the projection operator for an observable $A$ an observable and $a$ a subset of the values that this can take.
N.B. that this is in the ordinary-QM 2-time context, i.e. to be interpreted as {\sl subsequent} measurements.

\mbox{ }

\noindent {\bf Supplant at-a-time by value of a particular `clock' subconfiguration}.  
[Not all used of this necessarily carry `good clock' connotations: in the literature, this is done for whichever of the `any', `all' and `sufficient local' connotations.]
 
\mbox{ }  

\noindent {\bf Conditional Probabilities Interpretation (CPI)} \cite{PW83}.  
This goes further than the NSI by addressing conditioned questions of `being', and, moreover uses the Proposition--Projector Association in a {\sl non}standard context.  
Namely, the conditional probability of finding $B$ in the range $b$, given that $A$ lies in $a$, and to allot to it the value 
\beq
\mbox{Prob}(B\in b | A\in a; \Rho) = \frac{\mbox{tr}
\big(
\mP^B_{b}\,\mP^A_{a}\,\Rho\,\mP^A_a
\big)                            }{
\mbox{tr}
\big(
\mP^A_a\,\Rho
\big)
} \mbox{ } .   
\label{CPI}
\eeq
\noindent Note 1) It is these occurring within the one instant rather than ordered in time (one measurement and {\sl then} another measurement) 
that places this construct outside the conventional formalism of Quantum Theory, for all that (\ref{CPI}) superficially resembles (\ref{Pr:std}).

\noindent Note 2) The CPI can additionally deal with the question of `being at a time' by having the conditioning proposition refer to a `clock' subsystem \cite{PW83, I93}.  

\noindent Note 3) The CPI does implement propositions at the quantum level by use of projectors; its traditional development did not set up a scheme 
of logical propositions (it came historically before awareness of that began to enter the POT community via \cite{IL2, IL}).    
However, this layer of structure can be added to the CPI by the first principles reasons argued above.  

\noindent Note 4) {\bf Supplanting Questions of Becoming} in this kind of context has been considered e.g. in \cite{Page1, Page2, EOT, Records}.    
In Page's particular version, one does not have a sequence of a set of events; rather the scheme is a single instant that contains memories or other evidence of `past events'. 

\noindent Note 5) \K criticized the CPI in \cite{Kuchar92, Kuchar99} for its leading to incorrect forms for propagators.
Page answered \cite{PAOT} that this is a timeless conceptualization of the world, so it does not need 2-time entities such as propagators; see also 
Sec 14.  

\mbox{ } 

\noindent Records theory remains a heterogeneous subject, meaning that different people have postulated different axioms for it.


\noindent A) One can view the preceding extension as {\bf Page's form of Records Theory}.  

\noindent B) {\bf Bell--Barbour Records Theory} \cite{Bell,B94II,EOT}.  Take how $\alpha$-particle tracks form in a bubble chamber as a ``{\bf time capsules}" 
paradigm for Records Theory; perhaps Quantum Cosmology can be studied analogously \cite{B94II, EOT, H03, H09}, though here the paths are for the whole 
universe and in configuration space.  
Barbour's own approach has a number of additional elements \cite{B94II, EOT}; see \cite{FileR} for some commentary and caution.  

\noindent C) {\bf Gell-Mann--Hartle--Halliwell Records Theory} \cite{GMH} and Halliwell \cite{H99, HT02, H03} have found 
and studied records contained within Histories Theory (see the next SubSec).  

\noindent D) My own axiomatization for Records Theory prior to the above `semblance of dynamics' divergence is as follows \cite{Records}.  

\mbox{ } 

\noindent {\bf Records Postulate 1}. Records are information-containing subconfigurations of a single instant that are localized in both space and configuration space.
Local in configuration space concerns the imperfection of knowledge in practise, i.e. a notion of coarse graining.  
 
\noindent {\bf Records Postulate 2} Records can be tied to atemporal propositions, which, amount to a suitable logic.
\noindent [The tying at the quantum level is preferably by the Projector--Proposition Association. 
The form of the logical structure remains open to debate.  
The notion of localization in configuration space may well furnish the graining/partial order/logical implication operation.]  
\cite{ID, I10, FloriTh, F11} may be viewed as basis for implementation of a Records theory, though this is a bridge in the process of being built \cite{AF13}.  

\noindent {\bf Records Postulate 3}. Records are furthermore required to contain useful information. 
I take this to mean information that is firstly and straightforwardly about correlations. 
Secondly, however, one would wish for such correlation information to form a basis for a semblance of dynamics or history.  

\noindent Such a scheme therefore requires \cite{Records} suitable notions of locality in space and in configuration space, 
of propositional logic, of information, relative information and correlation.   

\noindent {\bf Records Postulate 4} \cite{ARel2} Semblance of time to be abstracted from records is to be of STLRC form. 
This is not as strong a selector as one might suspect, in that many timeless schemes built on Rovelli- or Barbour-type notions of 
Machian emergent time can be converted to this framework.  
Indeed, Records 1 to 4 can be seen as a new body in which to cast various approaches - a second-generation CPI/Page Records approach 
can have logic considerations and seek for conditioning to be on one's best estimate at that point on time abstracted from the STLRC.  
This is laid out in more detail in \cite{ARel2}.

\noindent {\bf Records Theory} is, finally, the subsequent study of how dynamics (or history or science) 
is to be abstracted from correlations between such same-instant subconfiguration records.

\subsection{Histories Theory}\label{Histor}

Perhaps instead it is the histories that are primary, a view brought to the GR context by Gell--Mann and, especially, Hartle  \cite{GMH, Hartle} 
and subsequently worked on by Isham, Linden, Savvidou and others \cite{IL2, IL, Savvidou02, Savvidou04, AS05, Kouletsis}.  

\mbox{ }  

\noindent {\bf Histories Postulate}. Treat histories (rather than configurations) as one's primary dynamical entities.

\mbox{ } 

\noindent Let us first assume that we treat histories at the classical level: Historia ante Quantum! ({\sc hq})
Then 

\noindent 
1) histories are sequences of configuration instants at given label-times, $\ttQ^{\sfA}(\tea_i)$ $i = 1$ to $n$ (discrete model) or curves in 
configuration space parametrized by a continuous label-time.  

\noindent 2) One's taking the histories to be the dynamical objects means that one has to define i) history momenta $\ttP_{\sfA}(\tea_i)$ that are canonically 
conjugate to the histories themselves in some new extended notion of phase space, and ii) history Poisson brackets in parallel to the usual Poisson brackets on the 
usual notion of phase space. 
Such approaches are then often referred to as `Histories Brackets' Approaches. 
iii) One then also has to define the corresponding histories quadratic constraint, ${\cal Q}$uad$\lfloor \ttQ^{\sfA}(\tea), \ttP_{\sfA}(\tea)\rfloor$ = 0. 

\noindent 
3) One also has notions of coarse- and fine-graining of histories (much as one does for records, and which in fact preceded that in the literature).   

\noindent Let us now conceive of histories at the quantum level, whether because one is doing `Historia Post Quantum' ({\sc qh}) or one is 
promoting the classical part of `Historia ante Quantum' ({\sc HQ}) to the quantum level.  

\noindent 0Q) Naively at the quantum level, one has a corresponding Feynman path integral structure (`sum over histories').  
However, Histories Theory itself is usually taken to have more structure than that.  

\noindent
1Q) Individual histories are now built as strings of projectors $\mP^{A_i}_{a_i}(\tea_i)$, i = 1 to $N$ at times $\tea_i$ (Hartle-type approach \cite{Hartle}), 
or a continuous limit of a tensor product counterpart (Isham--Linden {\bf Histories Projection Operator (HPO)}  Approach \cite{IL2}).
\noindent One distinction between these is that only the latter's products of projectors are themselves projectors and thus implementors of 
whole-histories propositions according to the very useful Proposition--Projector Association.  
Questions about histories are then another simplified form of logical structure as compared to temporal logic \cite{IL2, F10}.  
Note that the HPO Approach is a QFT {\sl in the (label) time direction}, even for `conventionally finite' models.  

\noindent 2Q) In the HPO approach, which is usually taken to be {\sc hq}, there is a kinematical commutator algebra of histories, and a quantum histories quadratic constraint.  

\noindent Regardless of whether it is an HPO Approach or of which {\sc h}, {\sc q} ordering is used, the following additional layers of structure are considered.

\noindent 3Q) There continue to be notions of coarse- and fine-graining at the quantum level.  

\noindent 4Q) Given a pair of histories $\gamma$, $\gamma^{\prime}$, the corresponding {\bf decoherence functional} is 
\beq
\mbox{Dec}(c_{\gamma^{\prime}}, c_{\gamma}) := \mbox{tr}(c_{\gamma^{\prime}}\Rho c_{\gamma}) \mbox{ } .
\eeq 
\noindent Note 1) In such as path integrals or composites thereof such as explicit computations of decoherence functionals, 
one is to use label time (or emergent time) when available (a few approaches use some kind of internal/matter time instead).  

\noindent Note 2) `Historia ante Quantum' ({\sc hq}) approaches can be seen as providing a second opportunity to a number of {\sc tq} 
approaches and group/geometrical quantization methods.  
E.g. new kinematical quantization, new commutator and constraint-bracket algebras/algebroids, and new possibilities for formulation of frame variables 
(such as Kouletsis' space maps \cite{Kouletsis}).

\noindent Note 3) Savvidou pointed out \cite{SavThes, AS05} that this version of Histories Theory has a distinct structure for each of two conceptually distinct notions of time: 

\noindent I) a kinematical notion of time that labels the histories as sequences of events (the `labelling parameter of temporal logic', 
taken by \cite{AS05} to mean causal ordering, though see also \cite{IL2}. 

\noindent II) A dynamical notion of time that is generated by the Hamiltonian.  

\noindent Savvidou has argued that having these two distinct notions of time allows for such a Histories Theory to be canonical and covariant at once, 
which is of obvious interest in understanding, and reconciling various viewpoints in, Quantum Gravity.  

\noindent Note 4) The records scheme sitting within a Histories Theory \cite{GMH, GMH11, H99, HT02, H03, HW06, H09, H11} is 
independent of the Gell-Mann--Hartle versus Isham--Linden distinction because these involve the single-time histories, 
i.e. a single projector, and then the ordinary and tensor products of a single projector obviously coincide and indeed trivially constitute a projector.  
Thus one can apply the Projector--Proposition Association and nicely found a propositional logic structure on this.

\mbox{ } 

\noindent It is convenient to end with some examples of (more or less) discrete approaches, 
some of which can be taken to be more structurally minimalist counterparts of Histories Theory.  

\noindent 1) {\bf Spin foams} \cite{Rovellibook, Thiemann, Baretal} can be viewed as path-integral counterparts of LQG.
As such, one might further build these approaches up to possess further histories-theoretic structure.
Schroeren's recent work has given a Hartle-type analysis for a spin foam \cite{Schroeren}.
Previously, Savvidou had cast Barbero-real-variables GR in HPO form, and some elements of Histories Theory 
also enters into composite attempts at resolving the POT in Ashtekar Variables programs, see e.g. \cite{Thiemann}.

\noindent 2) The {\bf Causal Sets Approach} \cite{Sorkin03, Surya} may be viewed as much like a Histories Theory in which less structure is assumed.  
Here, all that is kept at the primary level is the set of events and the causal ordering structure on these.  

\noindent 3) In juxtaposition and for later use, the {\bf Causal Dynamical Triangulation (CDT)} 
approach \cite{AL10} is a path integral/sum over histories approach but does not conventionally involve further histories-theoretic elements.
It does also retain causal structure, and considers as primary a continuum limit rather than about the discrete structure used to build up to that.
Thus it assumes somewhat more structure than in the Causal Sets Approach.

\section{Configurational relationalism generalization of Best Matching Problem}\label{CR}

The most general conceptual entity here is indeed Sec 1's Configurational Relationalism; a priori, there is no strong reason why this needs to be addressed 
(just) at the level of the classical Lagrangian formalism that is the domain of the Best Matching generalization of the Thin Sandwich Problem.
Nontrivial Configurational Relationalism produces linear constraints ${\cal L}$in$_{\sfZ}$, which need to be solved at {\sl some} level; 
up until that level, one may well need to consider other indirectly-formulated $G$-invariant objects (notions of distance, of information, quantum operators...). 
All of these and best-matched actions themselves are objects of the following $G$-act, $G$-all form\footnote{See \cite{FileR} for a more detailed 
account of this concept and its scope. 
It is a group sum/average/extremization type of move.}
e.g. the group action $\stackrel{\rightarrow}{G}_g$ followed by integration over all of the group $G$ itself, 
\beq
\mA^{G-\sf\sr\se\se} = \int_{g \in G} {\cal D}g \stackrel{\rightarrow}{G}_g \mA \mbox{ } \mbox{ for whichever kind of object A } . 
\eeq
This is limited for full quantum gravity by not being more than formally implementable for the case of the 3-diffeomorphisms.

Thus one needs to order not just {\sc t} and {\sc q} tuples, but also a {\sc r} = ({\sc r}, nothing) tuple.  
This generalizes `reduced quantization' and `Dirac quantization', which come from a context in which time-choosing was not necessary 
and simply mean {\sc r} ... {\sc q} and {\sc q} ... {\sc r}.
Moreover, this is but one of several further procedural ordering ambiguities.\footnote{See the next Sec for another, whilst {\sc q} itself has a 
fair amount of internal structure \cite{I84, ARel}: kinematical quantization, allotting a pre-Hilbert space, promoting constraints 
to operators and then solving them.  
Reduction itself can be split by solving for geometrically separable linear constraints at different points in the procedure. 
E.g. in RPM's, one can readily remove the translations at the start but quantize before constraining with respect to the rotations. 
Likewise, one might consider removing LQG's $SU(2)$ classically -- passing to Wilson loops -- prior to treating the Diff(3) constraint at 
the quantum level -- passing to quantum states depending only on spin-knots.}
%
Classically, there are a priori 8 schemes.  
This is the same combinatorial  8 as in Sec \ref{FFP}.  
See Appendix B for the countings. 
Including the quantum also, there are now 27 combinations of strategies.   
As further strategic diversity, one can also consider   `$\chi$ or plain' versions of {\sc t} and {\sc h} 
($\chi$ meaning time-and-frame constructions rather than the plain case's single-time constructions). 
However, the {\sc r} ... $\chi$ ordering makes no sense and so is to be discarded. 
On the other hand, {\sc tr} = {\sc rt} (at least at the conceptual level) since each of these procedures acts on separate parts of one's relational theory. 
Taking this further diversity and physical restriction into account too, there are 10 classical schemes (4 single, 4 frame and 2 timeless).
Allowing for the possibility of QM as well, there are now 35 schemes (16 single-time, 14 time-frame and 5 timeless).  

\mbox{ } 

\noindent Arena 1) Configurational relationalism is trivial in minisuperspace and whatever other ${\cal L}$in$_{\sfZ}$-bereft models.  

\noindent Arena 2) {\bf relational particle mechanics (RPM's)} \cite{BB82, FileR} are useful models for Configurational Relationalism, the Problem of Beables 
and midisuperspace nontrivialities.  
The notation below is $N$ particles in dimension $d$. 
$n = N - 1$ is used because of the triviality of removing translations.
Rotations Rot($d$) are here the analogues of Diff($\Sigma$), with a linear zero total angular momentum constraint $\mbox{\boldmath{${\cal L}$}}$ := $\sum_I \bq^I \cr \bp_I = 0$ 
corresponding to the GR momentum constraint and relationalspace R($N$, $d$) := $\mathbb{R}^{nd}$/Rot($d$) being the counterpart of superspace($\Sigma$).  
Dilations Dil as analogues of Conf($\Sigma$), corresponding to, in the pure-shape RPM case a ${\cal D} := \sum_I \bq^I \cdot \bp_I = 0$ 
constraint analogous to the GR maximum slicing condition, though in the scaled-RPM case ${\cal D}$ plays a similar role to the York time.
Here, the $I$ indexes particle labels, and $\bq^I$ are the particle positions with conjugate momenta $\bp_I$.  
Finally, preshape space is P($N$, $d$) := $\mathbb{R}^{nd}$/Dil = $\mathbb{S}^{nd - 1}$
    and     shape space is S($N$, $d$) := $\mathbb{R}^{nd}$/Rot($d$) \textcircled{S} Dil($d$), which is the counterpart of GR's CS($\Sigma$).  
I use $\mS^{\sfA}$ to denote shapes, $\sigma$ for scale (the total moment of inertia or its square root, in parallel to 
$\sqrt{h} = a^3$ and $a$ for cosmology), with corresponding shape momenta $p_{\sS}$ and scale momentum $p_{\sigma}$.
That particle models obviously have a notion of structure as well as of linear constraint, thus completing their midisuperspace-like features.
RPM's serve as toy models for JBB time, hidden time, semiclassical, records and histories approaches.  

\noindent Arena 3) In Geometrodynamics, resolving the Configurational Relationalism Problem at whatever level would entail working explicitly with the 3-geometries themselves 
from that point onward.  

\noindent Arena 4) As regards LQG, using loops, holonomies or spin-nets involve having taken out the $SU(2)$, whilst using knots entails the Diff(3) also being taken out.   

\mbox{ } 

\noindent Strategy 1) In the {\sc ct} scheme, the emergent JBB time now takes the Configurational Relationalism entwined form 
\beq
t^{\sJ\sB\sB}_{\sR\sP\sM} = t^{\sJ\sB\sB}_{\sR\sP\sM}(0)+\stackrel{\mbox{\scriptsize extremum $g \in G$}}
         {\mbox{\scriptsize of RPM action}} 
         \int ||\d_{g}\bQ||_{\sbm}/\sqrt{E - V}  \mbox{ } \mbox{ or } \mbox{ }   
 \mt^{\sJ\sB\sB}_{\sG\sR} = \mt^{\sJ\sB\sB}_{\sG\sR}(0)+\stackrel{\mbox{\scriptsize extremum $g \in$ Diff($\Sigma$)}}
         {\mbox{\scriptsize of GR relational action}}
         \int_{\Sigma}\d^3x \sqrt{h}\int||\d_{g}\bh||_{\mbox{\scriptsize \boldmath${\cal G}$}}/\sqrt{\mbox{Ric(\bx;\bh]}} 
\mbox{ } .
\eeq
[Here, $\d_g$ is the differential corrected by the infinitesimal group action of $G$, and $\bm$ is the mass matrix kinetic metric.]
Further features of this candidate timefunction are discussed in \cite{FileR, ARel2}.  
Moreover, the Best Matching Problem has been explicitly solved in 1- and 2-$d$ RPM's \cite{FORD, FileR} (for GR, this gives the 
Thin Sandwich Problem).    
Unfortunately while this {\sc ct} scheme yields a time function, it does not in any way unfreeze the GR Hamiltonian constraint or its energy constraint 
RPM analogue; thus a {\sc ctq} scheme based on it is not satisfactory.
One might thus use an internal or matter time instead, or postpone enquiring about a time that remains serviceable at the quantum level until one is at the quantum level.  

\noindent  Strategy 2) Moreover, there is also a direct approach for RPM's; this has no reduction step in use and yet the theory is nontrivial, via Kendall's 
procedure \cite{Kendall} for constructing the shape spaces for these cases at the metric level and my subsequent application of the Jacobi--Synge procedure 
that converts geometries to mechanical models \cite{FORD, FileR}.   
In this case the JBB time requires no group extremization, being formally the same structure as in Sec 3.

\noindent  Strategy 3) One could solve the linear constraints at the Hamiltonian level instead, e.g. in Geometrodynamics for the longitudinal potential $W^{k}$ part of $K_{ij}$.  
This gives e.g. the alternative formal {\sc ctq} scheme in which a single scalar $t$ is found by solving the Lichnerowicz--York equation with the 
value of $W^k$ found from solving the GR momentum constraint substituted back in.  
This follows from York's \cite{York72} work, which was indeed built as an alternative to Wheeler's original Thin Sandwich conceptualization.  
However, {\sc q} steps involving solutions of the Lichnerowicz-York equation  are particularly impassable.  
One might adhere to scale-and-frame auxiliary variables elimination from  constraints (and CMC slicing condition) in Lagrangian form, 
as arising e.g. from the action in \cite{ABFKO} or in connection with the distinct conformal thin sandwich reformulation \cite{CTS} of the GR initial value problem itself.  
I note that neither of these have been investigated in depth as p.d.e. systems in this particular regard of classical reduction at the Lagrangian 
level/actually solving rather than just posing the Best Matching Problem.
One could also follow solving the GR momentum constraint by finding a single `time-map' Histories Theory ({\sc chq} scheme).  

\noindent  Strategy 4) Either of the preceding {\sc c}-first moves could be followed up by a Tempus Nihil Est approach on the relational configuration 
space ({\sc cq}).  

\noindent  Strategy 5) One could find a time-frame function(al) (internal or from matter) and then classically reduce away the frame so as to pass to a 
new single time function(al) ($\chi${\sc cq}).  
However, as a second example of entwining of linear constraints \cite{Kuchar92}, the Internal Time approach's evolutionary 
canonical transformation's generating function needs to be a function of the initial and final slices' metrics in the classical 
configuration representation, implying a need for the {\sl prior} resolution of the Best Matching problem.
One could instead find a classical histories theory with `space-map' and `time-map' \cite{KK02, Kouletsis} , 
prior to reducing away the `space-map' structure to pass to a new `time-map' ($\chi_{\sH}${\sc cq})
The reduction here could e.g. be a `Best Matching of histories': solve the Lagrangian form of the linear histories constraints 
${\cal L}$in$\lfloor \ttQ^{\sfA}, \ttP_{\sfA}\rfloor$ = 0 for the histories auxiliary variables so as to remove these from the formalism.
It could however also be a reduction at the Hamiltonian level.

\noindent  Strategy 6) One could face the linear constraints after quantizing, for instance in a classical internal or matter time-and-frame finding {\sc tqr} 
scheme or in an emergent semiclassical time approach that has been enlarged to include h- and l-linear constraints ({\sc qrt} scheme \cite{Kieferbook, SemiclI, ACos2}.  

\noindent  Strategy 7) A Histories Theory may still have linear constraints at this stage ({\sc qr} ordering), 
in which case there is a nontrivial histories commutator constraint algebroid.  
If this is an {\sc hqr} ordering, the kinematical commutator algebra being obtained by selecting a subalgebra of the classical histories quantities, 
whose commutator suitably reflects global considerations, and the quantum histories constraints are some operator-ordering of their classical counterparts.
[This will not always form the same constraint algebroid that the histories Poisson brackets of the classical constraints did.]   

\noindent  Strategy 8) One could consider Records Theory with linear constraints and $G$-act, $G$-all constructed $G$-invariant notions of distance and of 
information \cite{FileR} and $G$-invariant operators at the QM level ({\sc qc}), or a Histories Theory whose decoherence functional is constructed in such 
a manner ({\sc qhc}).  

\noindent Thus, whilst a number of the above could be viewed as Best Matching Problem avoiding strategies, Configurational Relationalism itself is inevitable.

\mbox{ } 

\noindent{\bf Kucha\v{r}'s argument}.\footnote{This follows from my discussions with him in 2010.  
See the next Sec for its strongest connections with his written works.  
There is also support in Kuchar's writings for the reduced approach such as his emphasis on having resolved the 
Thin Sandwich Problem for the cylindrical-wave midisuperspace in \cite{K72}, and his use of the reduced approach for spherically-symmetric 
midisuperspace in \cite{K94}, though the latter work and others do also focus on Dirac quantization.}
%
The ordering in which the linear constraints are reduced out at the classical level, {\sc r} ... {\sc q}, is the physical ordering.  
This is in the context that, firstly, {\sc r} ... {\sc q} and {\sc q} ... {\sc r} do not always agree. 
Secondly, one would not expect that appending unphysical fields to the reduced description should change any of the physics of the of the true dynamical degrees of freedom.   
Thus, if they do differ, one should go along with the reduced version.  
I finally note that this argument logically extends to its histories counterpart:  {\sc h} ... {\sc q}.

\section{Strategizing about the Problem of Beables}\label{Beables}

I consider this to involve introduction of a fourth tuple {\sc o} whose options are, in decreasing order of stringency, 
({\sc d}, {\sc k}, {\sc p}, --) standing for (Dirac, Kucha\v{r}, partial, none).

\mbox{ } 

\noindent {\bf Dirac observables/beables} \cite{DeWitt67} alias {\bf constants of the motion} alias {\bf perennials} 
\cite{Kuchar93, +Perennials, KucharObs, Ear02, PG05, WuthrichTh, BF08} are any functionals of the canonical variables O = $\mD[ \ttQ^{\sfA}, \ttP_{\sfA} ]$ 
such that, at the classical level, their Poisson brackets with all the constraint functions ${\cal C}_{\sfX}$ vanish: 
\beq
\{ {\cal C}_{\sfX}, \mO\} = 0 \mbox{ }  
\label{DirObs}
\eeq
(perhaps weakly \cite{I93}).
Thus, e.g. for Geometrodynamics  
\beq
\{  {\cal H}_{\mu}, \mO  \} =  0 
\mbox{ } ,  \mbox{ } \mbox{ for } \mbox{ }                   
{\cal H}_{\mu} = [{\cal H}, {\cal H}_i] \mbox{ } .                     
\label{OHi0}
\eeq
Justification of the name `constants of the motion' conventionally follows from the total Hamiltonian being
$H\lfloor \Lambda^{\sfX} \rfloor = \int_{\sS}\d \mS \, \Lambda^{\sfX}{\cal C}_{\sfX}$ for multiplier coordinates $\Lambda^{\sfX}$, 
so that (\ref{DirObs}) implies 
\beq
\d{\mO}[\ttQ(\tea), \ttP(\tea)]/\d \tea = 0 \mbox{ } . 
\eeq  
\noindent {\bf True} \cite{Rov91} alias {\bf complete observables/beables} \cite{Rov02} (which at least Thiemann \cite{Thiemann} 
also calls evolving constant of the motion) are a similar notion, which, classically, involve operations on a system each of which produces 
a number that can be predicted if the state of the system is known.    

\mbox{ }  

Suppose one replaces (\ref{DirObs}) with split conditions
\beq
\{{\cal Q}\muu\ma\md, \mO\} = 0   
\mbox{ } , \mbox{ } \mbox{ } 
\{{\cal L}\mi\mn_{\sfZ}, \mO\} = 0  
\mbox{ } \mbox{ i.e. } \mbox{ } \mbox{ }    
\label{SplitCo} 
\{{\cal H}, O\} = 0 
\mbox{ } , \mbox{ } \mbox{ } 
\{{\cal H}_i, O\} = 0 \mbox{ for Geometrodynamics } .
\eeq
\noindent {\bf Kucha\v{r} observables/beables} \cite{Kuchar93} are then as above except that only the second bracket of (\ref{SplitCo}) need vanish. 
Kucha\v{r} then argued \cite{Kuchar93} for only the former needing to hold, in which case I denote the objects by K$[ \ttQ^{\sfA}, \ttP_{\sfA} ]$.
See also \cite{Kuchar93, KucharObs, Ear02, WuthrichTh, BF08, Kouletsis}.  

\mbox{ }  

\noindent Note 1) It is clear that finding these is a timeless pursuit: it involves configuration space or at most phase space but not the 
Hamiltonian constraint and thus no dynamics.
The downside now is that there is still a frozen ${\cal Q}$uad on the wavefunctions, so that one has to concoct some kind of 
Tempus Post Quantum or Tempus Nihil Est manoeuvre to deal with this.  

\noindent Note 2) If Configurational Relationalism has by this stage been resolved, knowledge of the \K observables is trivial. 
I.e. this is so any {\sc r} ... {\sc k} ordered approaches; thus there is a close link between Kucha\v{r}'s argument and making use of  
\K observables.

\noindent Example 1) (RPM's). In the pure-shape case, any functional F[S$^{\sfA}$, $\mP^{\sS}_{\sfA}$] = K, and in the scaled case, any functional 

\noindent F[S$^{\sfA}$, $\sigma$, $\mP^{\sS}_{\sfA}$, $\mP_{\sigma}$] = K; for 1- and 2-d RPM's a basis of shapes, a scale and the momenta conjugate to all 
of these are explicitly known \cite{QuadII}.

\noindent The next two examples are formal rather than explicitly constructed.

\noindent Example 2) Functionals of 3-geometries and conjugates are \K observables for Geometrodynamics.

\noindent Example 3) Functionals of knot configurations and conjugates are \K observables for LQG.  

\noindent Note 3) Formal $G$-act, $G$-all expressions for \K observables are also available if Configurational Relationalism has not by this stage been solved.

\mbox{ } 

The quantum counterpart of Dirac and \K beables involves some operator form for the canonical variables and quantum commutators 
$[ \mbox{ } , \mbox{ } ]$ in place of Poisson brackets.  
The operator-and-commutator counterparts of the Hamiltonian constraint constitute another `Heisenberg' manifestation of the FFP.
One's classical notion of observable is in each of the above cases to be replaced with the quantum one tied self-adjoint operators obeying 
a suitable commutator algebroid in place of the classical Poisson one; this correspondence is however nontrivial 
(e.g. the two algebroids may not be isomorphic) due to global considerations \cite{I84}.  
The Groenewold--Van Hove phenomenon is also an issue here if one tries to promote a classically-found set of observables to a QM set, i.e. in {\sc o} ... {\sc q} schemes.  

\mbox{ } 

\noindent {\bf S-matrix quantities} obey the first and not necessarily the second of (\ref{SplitCo}), or possibly the QM counterpart of this statement.  
These do not carry background-dependence connotations due to corresponding to scattering processes on configuration space rather than on space.
Clearly then $\mbox{Kucha\v{r} AND S-matrix } \Rightarrow \mbox{Dirac}$.
Moreover, Halliwell \cite{H03} supplies a {\sl specific construct} for a family of these, beginning from (for a simple particle mechanics model) 
\beq
A(\bq, \bq_0, \bp_0) = \int_{-\infty}^{+\infty} \d t \,  \delta^{(k)}(\bq - \bq^{\scc\sll}(t)) \mbox{ } , 
\mbox{ } 
{C}_{\sR} = \theta
 \left(   
\int_{-\infty}^{\infty} \d t f_{\sR}(\bq(t))  - \epsilon 
\right) P(\bq_{\sf}, \bq_0) \,   \mbox{exp}(iS(\bq_{\sf}, \bq_0)) \mbox{ } , \label{gyr}
\eeq
where the former is classical and the latter is a semiclassical QM {\it class function} (see \cite{HT02} for more details of this variant 
and Sec \ref{Combino} for discussion of more advanced variants).\footnote{Here, $\bq^{\scc\sll}(t)$ is the classical trajectory, 
$\bq_{0}, \bp_0$ is initial data, $\theta$ is the step function, $f_{\tR}$ is the characteristic function of region R, $\epsilon$ is a small 
number, $S(\bq_{\sf}, \bq_0)$ is the classical action between $\bq_{\sf}$ and $\bq_0$; see \cite{HT02} for the detailed form of  the prefactor function $P$.}   
  
It is important to treat the whole path rather than segments of it, since the endpoints of segments contribute right-hand-side terms to 
the attempted commutation with $H$ (Halliwell's specific context being that with no linear constraints ${\cal L}$in$_{\sfZ}$, so $\fH = \alpha{\cal H}$).  

\mbox{ }

\noindent Note 4) I remedied \cite{AHall} this no-${\cal L}$in$_{\sfZ}$ limitation using the RPM arena, in which I promoted the resolution Problem of Beables 
in the sense of \K to the sense of Dirac.     

\mbox{ } 

\noindent
{\bf Partial observables/beables}. The other type of timeless approaches (by Rovelli, Thiemann and Dittrich: see  \cite{Rovellibook, Thiemann, 
Dittrich, Dittrich2, APOT} for discussion and \cite{DeWitt67, PW83} for some earlier roots) are quite often used in further developments in LQG. 
These involve classical or QM operations on the system that produces a number that is possibly totally unpredictable even if the state is 
perfectly known (contrast with the definition of total/Dirac observables).
The physics then lies in considering pairs of these objects which between them do encode some extractable purely physical information.  
This is also an `any' move, in fact a second such: anything for time before, and anything to serve as partial beables.  
Here the Problem of Observables is held to have been a misunderstanding of the true nature of beables, which are in fact entities that 
are commonplace but meaningless other than {\sl as regards correlations between more than one such}.  

\noindent The {\sc p} (partial) option for beables can in principle be readily carried out at any stage relative to the other moves.  

\mbox{ }

\noindent {\bf Master constraint program observables/beables} can be conceived of, obeying not (\ref{DirObs}) but $\{\mO, \{\mO, \mathbb{M}\}\}$ = 0.
The conceptual meaning of these has not, as far as I am aware, been exposited.

\mbox{ }

\noindent {\bf Histories observables/beables} \cite{Kouletsis}. These are, conceptually, quantities that commute with the histories versions of the constraints.  
These are a distinct and meaningful concept in the Dirac and \K cases of observables/beables, whose definitions are constraint-dependent.  

\mbox{ } 

\noindent Note 5) The {\sc qort} classification, i.e. how to order {\sc q}, {\sc o}, {\sc r}, {\sc t} tuples in dealing with the quantum version of a theory with Background Independence.  
The relation-free count of number of possible orderings of maps is by now over 100 at the classical level and several times that 
for programs including quantization.  
These large numbers of programs illustrate the value of restricting these via firstly the commutations of some of the maps 
and secondly by further principles as in Note 7).
 
\noindent Note 6) Using the {\sc qort} classification to properly identify strategies.  
As an example, recognizing Thiemann's argument for `Dirac' rather than `reduced' \cite{T06} to involve only 
subsets of these.\footnote{\cite{Carlip01} 
also talks of clock fluctuations in this way, motivated e.g. by the status quo approach to path integrals for constrained theories.} 
The argument is based on extra freedom in clock choices in the Dirac picture has its fluctuations suppressed in the reduced case, rendering the reduced case less physical.  
However, this argument specifically involves favouring {\sc tqr} over {\sc trq} rather than for {\sc q} ... {\sc r} over {\sc r} ... {\sc q} in 
general that standard useage of `reduced quantization' would imply.  
Moreover, whilst the {\sc tqr} scheme having more variables means that it has a larger variety of such clock variables, these extra degrees of 
freedom are clearly unphysical and thus fluctuations in them are physically irrelevant, thus forming an effective counterargument.  
This alongside \K's argument makes for a strong case for {\sc r} ... {\sc q} approaches, contrary to much current literature.
The general suggestion is that gauge theory is unfortunately no longer physically viable in the presence of a ${\cal Q}$uad that corresponds to the 
Temporal Relationalism aspect of Background Independence.  
Thus a leading challenge for Quantum Gravity would appear to be how to redo classical gauge theory in gauge-invariant terms and follow through on quantizing that.
RPM's in 1- and 2-$d$ have the good fortune of being tractable in this way, but, for now, the extension of that to field theoretical models remains a major obstacle.  

\noindent Note 7) Furthermore, out of the remaining {\sc qort}'s, I argue in favour of {\sc rqot}, {\sc rqto}, {\sc rqoh}, {\sc rqho} and 
{\sc rqo} using {\sc k} or {\sc d} for the {\sc o}'s and noting that all the {\sc t}'s and {\sc h}'s are single times and not time-frames, and 
basing my choice on Note 6) Kucha\v{r}'s argument and the Machian Status Quo.  
Thus, provided that {\sc t}-{\sc h}-Nihil combinations are not being sought (for these it is equitable to set up {\sc t} and {\sc h} and the same 
relative position to {\sc q}, {\sc o}, {\sc r}), {\sc rhqo}, {\sc hrqo} and $\chi_{\sH}${\sc rqo} Histories Brackets approaches also fit these criteria.

\noindent Note 8) See Sec \ref{Global} for discussion of {\sl localized} notions of observables/beables.

\section{Strategizing about the Foliation Dependence Problem}\label{Fol}

\noindent In the continuum GR case, the $n^{\mu}$ of Sec 2 can furthermore be interpreted as the foliation 4-vector.
One should also ask what the analogues of spatial slices and of foliations are for discrete-type approaches to Quantum Gravity.
In CDT, spatial triangulations are the instants, and in spin foams, they are spatial spin network states.
In the Causal Sets Approach, this role is played by {\it antichains}; this makes good sense, since these are sets of causally {\sl unrelated} points.   
Moreover, since points and causal relations are {\sl all} of the structure in this approach, these antichains are just unstructured sets of points.
However, they can be slightly thickened so as to have enough relations to be structured.                                        
One can then imagine in each of these discrete approaches for layered structures of each's means of modelling an instant, and then pose questions of 
`foliation dependence' about this (or some suitable limit of this).  

\mbox{ } 

\noindent Attitude 1) Demand that classical GR's foliation independence and refoliation invariance continue to hold at the quantum level.  
One then has to face that the commutator algebroid of the quantum constraints is almost certainly distinct from 

\noindent classical GR's Dirac algebroid, 
and may contain foliation-dependent anomalies too.  
  
\noindent Attitude 2) Background-dependent/privileged slicing alternatives are approaches with times with sufficient significance imposed on them that 
they cannot be traded for other times (these are more like the ordinary Quantum Theory notion of time than the conventional lore of time in GR).  
As the below examples show, the amount of theorizing involving Attitude 2) has been on the increase; however, I first present a few cautions.

\mbox{ }

\noindent Note 1) One should disentangle Attitude 2) from how special highly-symmetric solutions {\sl can} have geometrically-preferred  foliations.
However, GR is about generic solutions, and even perturbations about highly-symmetric solutions cease to have geometrically-privileged foliations 
to the perturbative order of precision \cite{Kuchar99}.  
Additionally, even highly-symmetric solutions admitting a privileged foliation in GR are {\sl refoliable}, so the below problem with losing refoliation invariance does not apply.  

\noindent Note 2) As \K states \cite{Kuchar92}, {\it ``the foliation fixing prevents one from asking what would happen if one attempted to 
measure the gravitational degrees of freedom  on an arbitrary hypersurface. 
Such a solution 
amounts to conceding that one can quantize gravity only by giving up GR: to say that a quantum theory makes sense only when one fixes the foliation is 
essentially the same thing as saying that quantum gravity makes sense only in one coordinate system."}    

\noindent Note 3) On the other hand, one cannot press too hard with envisaging different foliations as corresponding to various motions of a 
cloud of observers distributed throughout space, since that could also be modelled by multiple congruences of curves without reference to multiple foliations.  


\noindent Example 2.1) Einstein--Aether theory \cite{AET} has privileged background structures due to a unit timelike vector field. 

\noindent Example 2.2) Shape Dynamics in the sense of \cite{GGKM11, K1108, GrybThesis} has a fixed CMC foliation. 
Here, one ``{\it trades refoliation invariance for a conformal symmetry}".
[Moreover, not everybody working in this area is holding out for a fixed-foliation interpretation.  
The situation is an enlarged phase space `linking theory', for which one gauge-fixing produces a GR sector and another gauge-fixing 
produces a CMC-fixed sector.
Then from the perspective of the linking theory, GR and CMC-fixed sector are gauge-related.]  

\noindent Example 2.3) Ho\v{r}ava--Lifshitz theory (HLT) \cite{Horava, Horava2} has a privileged foliation (which has also been identified \cite{Afshordi} as CMC).
\noindent In this approach, one gives objective existence to the foliation and then have solely the foliation-preserving subset of the 4-diffeomorphisms, Diff$_{\cal F}({\mM})$.
There are in fact various versions of HLT, corresponding to whether $\alpha = \alpha$(t alone) or $\alpha(\mt, \bx)$, crossed with 
what form the potential term is to take.  
The $\alpha = \alpha$(t alone) theories, by obvious parallel with Sec 3, avoid having a local ${\cal H}$ and thus FFP, 
though this is clearly at the expense of dropping many of the GR-like properties of time.
The other theories produce a local ${\cal H}$ and have been found to require extension in the number of included terms \cite{Healthy} so as to remain viable.

\noindent Example 2.4) Some of the more successful forms of CDT have been found to involve preferred foliations \cite{CDT2}. 

\noindent Example 2.5) Soo and Yu \cite{SY12} look to use a Master Constraint-type argument so as to also trade refoliation invariance for a privileged foliation theory.

\mbox{ }

\noindent Note 3) The above examples bear rich inter-relations.
If Einstein--Aether Theory's unit timelike vector is restricted to be hypersurface-orthogonal, the infra-red limit of the extended 
$\alpha(\mt, \bx)$ HLT is recovered \cite{AET2}. 
Searching for a suitable classical limit, the abovementioned CDT's produced HLT spacetime rather than GR spacetime \cite{CDT1, Horava2, CDT2}; 
Soo and Yu's approach  has links with HLT too.
Also see \cite{AHL} for further inter-relations between HLT and both the RWR and shape dynamics areas of Barbour's relational program.  

\noindent Note 4) Fixed foliations are a type of background-dependence, which, from a relational perspective, is undesirable {\sl from the outset}, 
for all that investigating the effects of making {\sl just} this concession is  indeed also of theoretical interest.

\mbox{ } 

\noindent As regards Attitude 1), it can be used to provide a strong basic counting argument against the unimodular approach.  
For, there can only be 1 degree of freedom in a time $t^{\su\sn\si}$ that arises from the cosmological constant, and this cannot 
possibly index the plethora of foliations corresponding to GR's many-fingered time notion.
The geometrical origin of this mismatch is that a cosmological time measures the 4-volume enclosed between two 
embeddings of the associated internal time functional $t^{\si\sn\st}$.  
However, given one of the embeddings the second is not uniquely determined by the value of $t^{\su\sn\si}$ 
(since pairs of embeddings that differ by a zero 4-volume are obviously possible due to the Lorentzian signature and cannot be distinguished in this way). 
Some constructive examples of Attitude 1) are as follows. 
 
\noindent Example 1.1) Kouletsis and \K \cite{KK02, Kouletsis} provided a means of including the set of foliations into an extension of the Arnowitt--Deser--Misner 
phase space that is generally covariant.
This amounts to extending phase space to include embeddings so as to take into account the discrepancy between the Diff({\cal M}) algebra and 
the Dirac algebroid (itself an older idea of Isham and Kucha\v{r}), here effectuated by construction of a time map and a space map.
Thus this is a $\chi_{\sH}$ {\sc k}... approach; it remains unstudied for GR past the classical level.
 
\noindent Example 1.2) In Savvidou's approach to HPO, she pointed out that the space of histories has implicit dependence on the foliation vector 
(the unit vector $n^{\mu}$ mentioned in Sec 2 as being orthogonal to a given hypersurface).
With this view of the HPO Approach admitting 2 types of time transformation, the histories brackets turn out to now be foliation-dependent. 
However, the probabilities that are the actual physical quantities, are {\sl not} foliation-dependent, so this approach avoids having a  
Foliation Dependence Problem at its end.

\noindent Example 1.3) Isham and Savvidou also considered the possibility of quantizing the classical foliation vector itself \cite{Savvidou02, IS01101}.  
Here,  

\noindent
\beq
\{\widehat{n}_{\mu}\Psi\}  = n_{\mu}\Psi \mbox{ } , \mbox{ } 
\{\widehat{p}^{\mu\nu} \Psi\} = i \left\{ n_{\mu}\frac{\pa}{\pa n_{\nu}} - n_{\mu}\frac{\pa}{\pa n_{\nu}}  \right\} \Psi \mbox{ } ,  
\eeq 
where the antisymmetric $p^{\mu\nu}$ is the conjugate of $n^{\mu}$ and satisfies the Lorentz algebra.
They then apply a group-theoretic quantization to the configuration space of all foliation vectors for the Minkowski spacetime toy model of the HPO Approach.

\section{Strategizing about the Functional Evolution Problem}\label{FEP}

Closure becomes complicated at the quantum level.
Firstly, one has to bear in mind that the classical and quantum algebras/algebroids are not in fact necessarily related due to global and Multiple Choice Problem 
considerations \cite{I84}.  

\noindent Suppose then that we experience non-closure.  

\noindent 1) One could try to blame this on operator-ordering ambiguities, and continue to try to obtain closure.

\noindent 2) Another counter that is sometimes available is to set some numerical factor of the obstructing term to zero; 
this is the same means by which String Theory arrives at its particular dimensionalities.

\noindent 3) Another possibility is to accept the QM-level loss of what had been a symmetry at the classical level. 

\noindent 4) Yet another possibility is that including additional matter fields could cancel off the anomaly hitherto found; 
some supergravity theories exhibit this feature. 

\noindent 5) A final possibility is to consider a new constraint algebroid, in particular a simpler one, along the lines of e.g. the 

\noindent $\alpha$($t$ alone) HLT or the Master Constraint Program.  
The latter cuts down the number of constraints so far that it would free the theory of all anomalies; however, that is in 
itself suspect since the same conceptual packaging of constraints into a Master Constraint would then appear to be applicable 
to the flat-spacetime gauge theory of standard Particle Physics, and yet these have not been declared to no longer have anomaly concerns.  

\mbox{ } 

\noindent On the other hand, other constraint algebroid variants enlarge the algebroid, e.g. more traditional forms of LQG algebroid
(anomaly analysis for which is in e.g. \cite{LQGAnomal}), histories algebroid \cite{Kouletsis} or a linking theory algebroid \cite{K1108}.  

\noindent Non-closure can be due to foliation-dependent terms, or, if scale is included among the physically irrelevant variables, 
then there can also be conformal anomalies; these are likely to plague Shape Dynamics.

\mbox{ }

\noindent Finally, note that the concept of `{\it observables/beables anomaly}', i.e. finding that promoting classical observables/beables 
to QM operators may produce quantities that do not commute with the quantum versions of the constraints. 
This will tend to add to the futility of {\sc o} ... {\sc q} schemes, since classical Problem of Beables solutions 
need not readily carry over to QM Problem of Beables solutions.  
At least in the simpler physical theories, \K observables/beables may well be largely exempt due to the nice properties of Lie groups 
under quantization schemes, but, in cases going beyond that (Dirac observables, the classical Dirac algebroid...) one may find one has 
to solve the QM Problem of Beables again. 
See the Halliwell combined scheme of Sec 14.1 for a good example, there are separate classical and QM implementations for Dirac-type observables/beables, 
whilst the \K observables translate over straightforwardly for the triangle RPM model \cite{AHall}.

\section{Strategizing about the Global Problem of Time}\label{Global}

\noindent Global Attitude 1) insist on only using globally-valid quantities for timefunctions and observables.

\mbox{ } 

\noindent Example 1.1 Scale times are not useable globally-in-time in recollapsing universes due to non-monotonicity. 
Passing to dilational times (conjugate to scales) such as York time remedies this problem. 

\noindent Example 1.2) The global POT then occurs e.g. in the separation into true and embedding (space frame and timefunction) 
variables in the internal time approach (the Torre Impasse \cite{Torre}).  

\noindent Example 1.3) Moreover the above example's split might be defineable in some case, but only for some sequence of slices 
that cannot be indefinitely continued as one progresses along a foliation that is to cover the entirety of that spacetime 
(e.g. not all spacetimes can be entirely foliated by CMC slices).  

\noindent Example 1.4) Simpler models' timefunctions can eventually going astray for some other reason not involving foliations 
[after all, many simpler models do not have (meaningful) notions of foliation].  
E.g. in the semiclassical approach, the allotment of h-l status may be only local in configuration space, with the approximate 
dynamical trajectory/wavepacket free to leave that region, and likewise the WKB regime itself may only be local, and all of this 
can already happen e.g. in RPM's for which the notion of foliation is moot.    

\noindent Example 1.5) The Global POT can partly be avoided by using some timeless approach, though elements of this construct including the 
semblance of dynamics itself may be local.

\mbox{ }  

\noindent Global Attitude 2) allow for use of patching of various notions of time and of observables that are only locally valid \cite{Benito, Bojo}. 
Whilst this was very naturally covered at the classical level by time being a coordinate and thus only in general locally valid and 
subject to the ordinary differential-geometric meshing condition, the problem now is how to mesh together different unitary evolutions.  
Many of the other POT facets themselves have global issues (e.g. global best-matching resolution, global observables, globally valid foliations, globally valid 
spacetime reconstructions). 

\mbox{ } 

\noindent Example 2.1) Bojowald et al's {\bf fashionables} concept \cite{Bojo}, or my {\bf  degradeables} parallel \cite{FileR, AHall}. 
Fashionables are observables local in time and space, whereas degradeables are beables that are local in time and space.  
These are good words for local concepts, viz `fashionable in Italy', `fashionable in the 1960's', `degradeable outside of the freezer' 
and `degradeable within a year' all making good sense. 
Also, fashion is in the eye of the beholder -- observer-tied, whereas degradeability is a mere matter of being, rather than of any observing.  
These are {\bf patching approaches}: observables/beables, and timefunctions, are held to only be valid on certain local patches.  
This holds for any of the Dirac, \K and partial variants. 

\noindent Example 2.2) Bojowald et al's patching approach uses a moments expansion as a bypass on the inner product problem.
It is also semiclassical, in their sense that they neglect O($\hbar^2$) and moment-polynomials above some degree.  
Bojowald et al's specific fashionables implementation is based on time going complex around turning points within their notion of 
semiclassical regime. 
However, the geometrical interpretation of this transition between these fashionables is, for now, at least to me, unclear.  

\noindent Example 2.3) At the level of \K observables for RPM's, using fashionables/degradeables means that the functionals of the shapes 
(and scales) and their conjugates do not need to be valid over the whole of configuration space. 
Sec 8's examples can be upgraded likewise to be about functionals that are local in the corresponding configuration spaces.  
RPM's should also allow for Bojowald's working to be considered in cases which posses the major midisuperspace features.

\section{Strategizing about the Multiple Choice Problem}\label{MC}

Whilst patchings are multiple, and \cite{Bojo} claim that as an Multiple Choice Problem solver too, this does not address whole of this problem.  
Generally, the Multiple Choice Problem is multi-faceted -- due to a heterogeneous collection of mathematical causes.    
Multiplicity of times is also inherent in Rovelli's `any change' and `partial observables' based relationalism, and in my STLRC 
(where one tests one's way among the many to find the locally-best).

\mbox{ }

\noindent 1) Some cases of Multiple Choice Problem reflect foliation dependence.  
It has been suggested (e.g. in \cite{unim-SUGRA}) that one way out of the Multiple Choice Problem is to specify the lapse $\alpha$ and shift $\beta^{i}$ a priori . 
However, such amounts to yet another case of 

\noindent foliation fixing (so return to Sec \ref{Fol}).   

\noindent 2) It may be that some cases of Multiple Choice Problem could be due to context: different observers observing different subsystems that 
have different notions of time; this fits in well with the partial observables and patching paradigms.  

\noindent 3) Some cases of Multiple Choice Problem involve how choosing different times at the classical level for subsequent use (`promotion to operators') 
at the quantum level can lead to unitarily-inequivalent QM's, even if at the classical level they are canonically-related.  
This is due to the Groenewold--Van Hove phenomenon (which already holds for finite theories).
This is a time problem if one's search for time takes one among the classical observables (i.e. {\sc o ... t ... q} approaches). 

\noindent 4) Possible problems with Patching are that the moments approach looks to rely on classical and quantum having the same kinematical bracket algebra (generally untrue \cite{I84}).  
The moments approach's use of all polynomials (up to a given degree in the approximate case detailed in \cite{Bojo}) may be at 
odds with the Groenewold--Van Hove phenomenon, indicating that subalgebra-selection reasons for the Multiple Choice Problem are not included.  
Thus only a partial resolution of the Multiple Choice Problem looks to be on offer here.
The Groenewold--Van Hove part of the Multiple Choice Problem is at least part-avoidable via the semiclassical approach being less reliant on kinematical subalgebra 
choices or unitary inequivalences.

\section{Strategizing about the Spacetime Reconstruction Problem}\label{SRP}

\noindent A further Spacetime Reconstruction Problem Subfacet is that, in Internal Time Approaches, internal space or time coordinates are to be used 
in the conventional classical spacetime context; these need to be scalar field functions on the spacetime 4-manifold.    
In particular, functions of this form do not have any foliation dependence.
However, the canonical approach to GR uses functionals of the canonical variables, and which there is no a priori reason for such to be scalar fields of this type.  
Thus one is faced with either finding functionals with this property (establishing Foliation Independence by construction and standard spacetime interpretation recovery), 
or with coming up with some new means of arriving at the standard spacetime meaning at the classical level.   

\noindent Microcausality recovery is possible in the Savvidou \cite{Savvidou04} or Kouletsis \cite{Kouletsis} canonical-covariant formalisms
(though this work has as yet very largely not been extended to the quantum level).  

\noindent Programs are indeed often designed so that the continuum with spacetime properties recovery is the last facet to face.  

\mbox{ } 

\noindent Example 1) The status of the Spacetime Reconstruction Problem remains unclear for the Semiclassical Approach once examined in detail \cite{Kuchar92, I93}.  

\noindent The next examples involve bottom-up approaches, i.e. less structure assumed.  

\noindent Example 2) Getting back a semblance of dynamics or a notion of history from Timeless approaches counts as a type of Spacetime Reconstruction. 
This is in contradistinction to approaches in which histories are assumed.

\noindent Example 3) In LQG, semiclassical weave states have been considered by e.g. Ashtekar, Rovelli, Smolin, Arnsdorf and Bombelli; 
see \cite{Thiemann} for a brief review and critique of these works. 
Subsequent LQG semiclassical reconstruction work has mostly involved instead (a proposed counterpart of the notion of) 
{\it coherent states} constructed by the complexifier method \cite{Thiemann}; see \cite{FL11} for a different take on such.  
In Lorentzian spin foams, semiclassical limits remain a largely open problem; e.g. \cite{Baretal} is a treatment of Lorentzian spin foams that 
also covers how the Regge action emerges as a semiclassical limit in the Euclidean case, whilst almost all the 
semiclassical treatment in the most recent review \cite{Perez} remains Euclidean.  
The further LQC truncation \cite{Bojowald05} does possess solutions that look classical at later times (amidst larger numbers of solutions 
that do not, which are, for now, discarded due to not looking classical at later times, which is somewhat unsatisfactory in replacing 
predictivity by what amounts to a {\sl future} boundary condition). 
It also possesses further features of a semiclassical limit \cite{Bojowald06} (here meaning a WKB regime with powers of both $\hbar$ and 
Immirzi's $\gamma$ neglected, whilst still subject to open questions about correct expectation values of operators in semiclassical states).

\noindent Example 4) CDT succeeds in generating a classical regime \cite{AL10}, albeit this is HLT rather than GR, which is subject to 
the various foliation-related issues in Sec \ref{Fol} as well as non-POT related advantages in ultra-violet behaviour.    
From this, we may deduce that sometimes reconstructing spacetime is more straightforward if we forfeit its refoliation-invariance property 
(i.e. trading completion of one facet for taking the consequences of refusing to face another).  
See also \cite{SemiclCDT} for a semiclassical analysis of CDT.  

\noindent Example 5) Spacetime Reconstruction difficulties are to be expected from the Causal Sets Approach's insistence on very sparse structure;  
for recent advances here, see e.g. \cite{MRS} for a recovery of a spacetime-like notion of topology, or \cite{RiWa} for a recovery of a metric notion.   

\noindent Example 6) Euclidean counterparts of dynamical triangulation run into Wick rotation ambiguities (a similar situation to that observed 
in \cite{HL90}), and \cite{AL10} argue this to be evidence for the need to maintain the notion of causality.

\section{Combined strategies for the POT}\label{Combino}

I favour \cite{FileR, ARel, ARel2, ACos2} the Machian $\tea^{\sJ\sB\sB}$ as classical-level resolution of the POT modulo solving the 
Best Matching Problem.
One is then to continue at the quantum level by replacing $\tea^{\sJ\sB\sB}$ by the equally-Machian $\tea^{\sW\sK\sB}$.  
One is to justify the underlying WKB ansatz by some combined approach, the most promising of which is along the lines of Halliwell's 
histories, records and semiclassical combined approach (see Sec 14.1).  
It may be possible to use a more minimalistic combined scheme instead; the final Subsection 14.2 contains two such.
As regards these schemes involving semiclassicality, whilst this is a limitation for other purposes, is argued to be valuable e.g. in the 
quantum-cosmological setting in footnote \ref{gnimmel}.

\subsection{Halliwell-type combined strategies}

\noindent Both histories and timeless approaches lie on the common ground of atemporal logic structures \cite{IL, FileR}, and there 
is a Records Theory within Histories Theory, as per Sec 6.4. 
The Semiclassical Approach and/or Histories Theory could support Records Theory by providing a mechanism for the semblance of dynamics.    
Histories decohereing is a leading (but as yet {\sl not fully established}) way by which the semiclassical regime's WKB approximation could be 
legitimately obtained in the first place.  
The elusive question of which degrees of freedom decohere which should be answerable through where in the universe the information is actually 
stored, i.e. where the records thus formed are \cite{GMH, H03}; as Gell-Mann and Hartle say \cite{GMH},
\noindent 
\beq
\mbox{Records are ``{\it somewhere in the universe where information is stored when histories decohere}"}. 
\label{situ}
\eeq
Emergent semiclassical time amounts to an approximate semiclassical recovery \cite{SemiclI, SemiclIII, ACos2} of emergent classical JBB time \cite{B94I} is an encouraging 
result as regards making such a Semiclassical--Timeless Records combination.

More concretely, the semiclassical approach aids in the computation of timeless probabilities of histories entering given configuration space regions.
This is by the WKB assumption giving a semiclassical flux into each region \cite{H03}; this approach does not, however, proceed via 
an emergent semiclassical time dependent Schr\"{o}dinger equation such as in Sec 6.2.   
These timeless probabilities, and decoherence functionals, can be built from the class-function objects of eq (36 ii). 
These are additionally S-matrix quantities, resolving part of the Problem of Beables, and a first simple model (RPM) with linear constraints 
included has also been resolved \cite{AHall}.  
Finally these are indeed locally-interpretable concepts and so are furthermore compatible with the basic conceptual ethos of fashionables/degradeables, 
giving some hope of global POT resolution and partial Multiple Choice resolution, so this approach comes 
fairly close to resolving the POT in semiclassical regimes, and {\sl very} close to resolving the POT for RPM's themselves.

Moreover, whilst conceptually illustrative, the class function (36 ii) itself has technical problems, in particular it suffers 
from the Quantum Zeno problem. 
This can be dealt with firstly by considering the following reformulation \cite{HW06}:
\beq
\mP(N\epsilon) ... \mP(2\epsilon)\mP(\epsilon) = \mbox{exp}(-iHt) \mbox{ } ,
\eeq 
as well as now conceptualizing in terms of probabilities of {\sl never} entering the region.  
Next, one applies Halliwell's `softening' (in the sense of scattering theory), passing to 
\beq
\mP(N\epsilon) ... \mP(2\epsilon)\mP(\epsilon) = \mbox{exp}(-i(H - iV_0 P)t)
\eeq
for $\epsilon V_0 \approx 1$.  
The class operator for {\sl not} passing through region R is then
\beq
C_{\overline{\sR}} \mbox{ } \mbox{ } = \mbox{ } \mbox{ } \stackrel{\mbox{lim}}{\mbox{\scriptsize{$t_1 \longrightarrow -\infty, t_2 \longrightarrow \infty$}}} 
\mbox{exp}(i H t_2) \mbox{exp}(-i\{H - iV\}\{t_2 - t_1\}) \mbox{exp}(- i H t_1)
\mbox{ } .  
\eeq
This now does not suffer from the Quantum Zeno Problem whilst remaining an S-matrix quantity and thus commuting with ${\cal Q}$uad.

\subsection{CPI--semiclassical combined strategies}

The most developed of these to date is Gambini--Porto--Pullin's \cite{PGP1, PGP2}.  
This is built upon a different formulation of conditional probabilities objects: 
\beq
\mbox{\scriptsize Prob(observable ${O}$ lies in interval 
$\Delta{O}$ provided that clock variable $\tea$ lies in interval $\Delta\tea$)} = 
\stackrel{\mbox{lim}}{T \longrightarrow 0}
\frac{    \int_0^T\d \tea \mbox{Tr}(    \mP_{\Delta {O}}(\tea)  \mP_{\Delta \stea}(\tea) \Rho_0 \mP_{\Delta \stea}(\tea)  )}
     {    \int_0^T\d \tea \mbox{Tr}(\Rho_0 \mP_{\Delta \stea}(\tea))}  \mbox{ } .
     \label{NewP}
\eeq
The P($\tea$)'s here are Heisenberg time evolutions of projectors P: $\mP(\tea) = \mbox{exp}(iH\tea)\,\mP\,\mbox{exp}(-iH\tea)$
These conditional probabilities allow for `being at a time' to be incorporated.  
This approach is for now based on the `any change' notion of relational clocks (that can be changed \cite{ARel2} to the STLRC notion).
Moreover, these clocks are taken to be non-ideal at the QM level, giving rise to 

\noindent 1) a decoherence mechanism (distinct from that in Histories Theory; there is also a non-ideal rod source of decoherence postulated; 
the general form of the argument is in terms of imprecise knowledge due to imprecise clocks and rods). 

\noindent 2) A modified version of the Heisenberg equations of motion of the Lindblad type, taking the semiclassical form 
\beq
i\hbar\frac{\pa \Rho}{\pa \tea} = [H, \Rho] + D[\Rho] \mbox{ } .
\label{Lindblad}
\eeq
The form of the $D$-term here is $\sigma(\tea)[H, [H, \Rho]]$  where $\sigma(\tea)$ is dominated by the rate of change of width of the probability distribution.
By the presence of this `emergent becoming' equation (\ref{Lindblad}), this approach looks to be more promising in practise than Page's extended form of CPI. 
This approach also has the advantage over the original CPI of producing consistent propagators.  
By the presence of the $D$-term, the evolution is not unitary, which is all right insofar as it represents a system with imprecise knowledge (c.f. Sec 6.2).    
Dirac observables for this scheme are furthermore considered in \cite{GPPT}.  

\mbox{ } 

\noindent Note 1) One advantage of this scheme over the Halliwell-type one is that it is based on the projector-proposition 
implementation rather than a classical regions implementation.

\noindent Note 2) Arce \cite{Arce} has provided a distinct combined CPI--semiclassical scheme [for yet another notion of semiclassicality, 
that keeps some diabatic terms and constitutes an example of self-consistent approach, c.f. Sec 6.2].  

\mbox{ } 

\noindent {\bf Acknowledgements}: I thank: my wife Claire, Amelia, Sophie, Sophie, Anya, Amy, Hettie, Hannah, Becky, Lynnette, Fletcher, Sharon and Zander 
for being supportive of me whilst this work was done.   
Professors Julian Barbour, Jeremy Butterfield, Fay Dowker, Jonathan Halliwell, Petr Ho\v{r}ava, Chris Isham, Karel Kucha\v{r}, Daniele Oriti,  Jorge Pullin, 
Carlo Rovelli and Gerard t'Hooft, and Dr's Cecilia Flori, Sean Gryb, Tim Koslowski and James Yearsley, and Mr's Pui Ip, Matteo Lostaglio, David Schroeren and 
one anonymous Referee for discussions between 2008 and the present, and Professor Claus Kiefer for recommending me as a reviewer for this topic.  
I thank Profs Marc Lachi$\grave{\me}$ze-Rey, Pierre Binetruy and Jeremy Butterfield for helping with my career.    
I am funded by a grant from the Foundational Questions Institute (FQXi) Fund, a donor-advised fund of the Silicon Valley Community Foundation on 
the basis of proposal FQXi-RFP3-1101 to the FQXi.  
I thank also Theiss Research and the CNRS for administering this grant.  
I also thank DAMTP Cambridge and the Perimeter Institute Waterloo Canada for hospitality.

\mbox{ }  

\noindent{\bf \large Appendix A: Acronyms used in this Review} 

\mbox{ }

\noindent CDT \mbox{ } \mbox{ } \mbox{ }                              Causal Dynamical Triangulation: an approach to Quantum Gravity \cite{CDT1}.

\noindent CMC \mbox{ } \mbox{ } \mbox{ }                              constant mean curvature: a useful type of spatial slicing in GR.

\noindent CPI \mbox{ } \mbox{ } \mbox{ }    \,                        \CPI of QM: a timeless approach to the Problem of Time.

\noindent FFP \mbox{ } \mbox{ } \mbox{ }    \hspace{0.01in}           Frozen Formalism Problem: a facet of the Problem of Time. 

\noindent GR  \mbox{ } \mbox{ } \mbox{ } \, \hspace{0.015in}          General Relativity.

\noindent HLT \mbox{ } \mbox{ } \mbox{ } \hspace{0.01in}              Ho\v{r}ava--Lifshitz Theory: another approach to Quantum Gravity \cite{Horava}.  

\noindent HPO \mbox{ } \mbox{ } \mbox{ }    \hspace{0.005in}                           Histories Projection Operator: a type of Histories Theory \cite{IL}.

\noindent JBB \mbox{ } \mbox{ } \mbox{ }    \hspace{0.02in}           Jacobi--Barbour--Bertotti: a type of emergent time.

\noindent LQC \mbox{ } \mbox{ } \mbox{ }                              Loop Quantum Cosmology \cite{Bojowald05}.

\noindent LQG \mbox{ } \mbox{ } \mbox{ }                              Loop Quantum Gravity   \cite{Thiemann}.

\noindent NSI \mbox{ } \mbox{ } \mbox{ }  \,                          \NSI of QM: another timeless approach to the Problem of Time.

\noindent POT \mbox{ } \mbox{ } \mbox{ }                              Problem of Time.

\noindent QFT \mbox{ } \mbox{ } \mbox{ }                              Quantum Field Theory.

\noindent QM  \mbox{ } \mbox{ } \mbox{ } \hspace{0.01in}                            Quantum Mechanics. 

\noindent RPM \mbox{ } \mbox{ } \mbox{ }                              relational particle mechanics: a useful toy model for studying the Problem of Time \cite{FileR}.

\noindent RWR \mbox{ } \mbox{ } \mbox{ }                              relativity without relativity: a relational/Machian recovery of GR \cite{RWR}.

\noindent SR  \mbox{ } \mbox{ } \mbox{ } \,\,\,  \hspace{0.01in}                        Special Relativity.

\noindent STLRC \mbox{ }                 \,                             sufficient totality of locally-relevant change: a relational/Machian term \cite{ARel2}.

\noindent WDE \mbox{ } \mbox{ } \mbox{ }                              Wheeler--DeWitt equation: the QM wave equation for GR.

\noindent WKB \mbox{ } \mbox{ } \mbox{ }                              Wentzel--Kramers--Brillouin: a semiclassical approximation method, or the corresponding emergent time. 


\noindent{\bf \large Appendix B: Counting numbers of maps} 

\mbox{ } 

\noindent We want to know all possible compositions of maps where each map is used once or not at all. 
There are $K(n):= \sum_{n}P(n, r)$ of these for $P(n,r)$ the permutator, and this sum is computationally equal to int(e\,$n$!) for 
$n\geq 1$. 
The first few values of this are $K(1) = 2$, $K(2) = 5$, $K(3) = 16$, $K(4) = 65$...
Next, when one of the maps is chosen from a menu of two ({\sc t}, {\sc h} or nothing), the total number of maps is $D(n) := 2K(n) - K(n - 1)$.
The first few values of this are $D(2) = 8$, $D(3) = 27$, $D(4) = 114$...
Further counting in this article just involve i) more/larger menus: choosing 1 or none from each but also having the courses in any order.
ii) Relation restrictions on the maps: some maps are disallowed to the right of others, and some adjacent pairs of maps are equivalent by commutation.  
These follow from the more detailed mathematical meaning of the maps in question.


\end{document}